\documentclass[useAMS,usenatbib]{mn2e}

\usepackage{graphicx}
\usepackage{amssymb}
\usepackage{amsmath}

\newcommand{\diff}{\ensuremath{{\rm d}}}
\newcommand{\ii}{\ensuremath{{\rm i}}}
\newcommand{\ee}{\ensuremath{{\rm e}}}

\newcommand{\Herm}{\ensuremath{H_{m}}}
\newcommand{\Hermz}{\ensuremath{H_{0}}}

\newcommand{\bB}{\ensuremath{\mathbf{B}}}
\newcommand{\bv}{\ensuremath{\mathbf{v}}}

\newcommand{\bgrad}{\ensuremath{\mathbf{g}_{\rm rad}}}
\newcommand{\bgeff}{\ensuremath{\mathbf{g}_{\rm eff}}}

\newcommand{\deltav}{\ensuremath{\delta_{v}}}

\newcommand{\kappae}{\ensuremath{\kappa_{\rm e}}}
\newcommand{\Gammae}{\ensuremath{\Gamma_{\rm e}}}
\newcommand{\matom}{\ensuremath{u}}

\newcommand{\Qbar}{\ensuremath{\bar{Q}}}

\newcommand{\cool}{\ensuremath{\Lambda}}
\newcommand{\coolat}{\ensuremath{\cool_{\rm at}}}
\newcommand{\coolic}{\ensuremath{\cool_{\rm ic}}}

\newcommand{\coolcoeff}{\ensuremath{\mathcal{L}}}

\newcommand{\Uph}{\ensuremath{U_{\rm ph}}}

\newcommand{\rrot}{\ensuremath{r}}
\newcommand{\trot}{\ensuremath{\theta}}
\newcommand{\prot}{\ensuremath{\phi}}

\newcommand{\rmag}{\ensuremath{\tilde{r}}}
\newcommand{\tmag}{\ensuremath{\tilde{\theta}}}
\newcommand{\pmag}{\ensuremath{\tilde{\phi}}}

\newcommand{\xmag}{\ensuremath{\tilde{x}}}
\newcommand{\ymag}{\ensuremath{\tilde{y}}}
\newcommand{\zmag}{\ensuremath{\tilde{z}}}

\newcommand{\ximg}{\ensuremath{x_{\rm i}}}
\newcommand{\yimg}{\ensuremath{y_{\rm i}}}
\newcommand{\zimg}{\ensuremath{z_{\rm i}}}

\newcommand{\Aeq}{\ensuremath{A_{\rm eq}}}

\newcommand{\sstarN}{\ensuremath{s_{\rm N}}}
\newcommand{\sstarS}{\ensuremath{s_{\rm S}}}

\newcommand{\tmagN}{\ensuremath{\tmag_{\rm N}}}
\newcommand{\tmagS}{\ensuremath{\tmag_{\rm S}}}

\newcommand{\Bz}{\ensuremath{B_{0}}}
\newcommand{\Beq}{\ensuremath{B_{\rm eq}}}

\newcommand{\drot}{\ensuremath{\varpi}}

\newcommand{\berrot}{\ensuremath{\mathbf{e}_{\rrot}}}

\newcommand{\bermag}{\ensuremath{\mathbf{e}_{\rmag}}}
\newcommand{\betmag}{\ensuremath{\mathbf{e}_{\tmag}}}

\newcommand{\bes}{\ensuremath{\mathbf{e}_{s}}}
\newcommand{\bedrot}{\ensuremath{\mathbf{e}_{\drot}}}

\newcommand{\rKep}{\ensuremath{r_{\rm K}}}

\newcommand{\rAlf}{\ensuremath{r_{\rm A}}}

\newcommand{\smin}{\ensuremath{s_{0}}}
\newcommand{\hmin}{\ensuremath{h_{0}}}
\newcommand{\rhomin}{\ensuremath{\rho_{0}}}
\newcommand{\hinf}{\ensuremath{h_{\infty}}}

\newcommand{\poteff}{\ensuremath{\Phi_{\rm eff}}}
\newcommand{\poteffmin}{\ensuremath{\Phi_{0}}}

\newcommand{\rhopert}{\ensuremath{\rho_{\rm p}}}
\newcommand{\ppert}{\ensuremath{p_{\rm p}}}
\newcommand{\vpert}{\ensuremath{v_{\rm p}}}

\newcommand{\estar}{\ensuremath{\eta_{\ast}}}

\newcommand{\Nsurf}{\ensuremath{N_{\rm s}}}
\newcommand{\Nmag}{\ensuremath{N_{\rm m}}}
\newcommand{\deldisk}{\ensuremath{\Delta_{\rm d}}}
\newcommand{\delstar}{\ensuremath{\Delta_{\ast}}}

\newcommand{\hdisk}{\ensuremath{h_{\rm d}}}
\newcommand{\sigdisk}{\ensuremath{\sigma_{\rm d}}}
\newcommand{\sdisk}{\ensuremath{s_{\rm d}}}
\newcommand{\sigdiskrfhd}{\ensuremath{\sigma_{\rm d, RFHD}}}
\newcommand{\sigdiskrrm}{\ensuremath{\sigma_{\rm d, RRM}}}

\newcommand{\tmagdisk}{\ensuremath{\tmag_{\rm d}}}
\newcommand{\tmagdiskrfhd}{\ensuremath{\tmag_{\rm d, RFHD}}}
\newcommand{\tmagdiskrrm}{\ensuremath{\tmag_{\rm d, RRM}}}

\newcommand{\rhostar}{\ensuremath{\rho_{\ast}}}
\newcommand{\pstar}{\ensuremath{p_{\ast}}}
\newcommand{\rhosonic}{\ensuremath{\rho_{\rm s}}}
\newcommand{\csoundstar}{\ensuremath{a_{\ast}}}

\newcommand{\sigdot}{\ensuremath{\dot{\sigma}_{\rm d}}}

\newcommand{\Mdot}{\ensuremath{\dot{M}}}
\newcommand{\vinf}{\ensuremath{v_{\infty}}}

\newcommand{\Rsigma}{R(\sigdisk)}
\newcommand{\Dtmag}{\Delta(\tmag_{\rm d})}

\newcommand{\np}{\ensuremath{n_{\rm p}}}
\newcommand{\nel}{\ensuremath{n_{\rm e}}}
\newcommand{\Np}{\ensuremath{N_{\rm p}}}

\newcommand{\Tlo}{\ensuremath{T_{\rm 1}}}
\newcommand{\Thi}{\ensuremath{T_{\rm 2}}}

\newcommand{\emd}{\ensuremath{\mathcal{E}}}
\newcommand{\dem}{\ensuremath{\mathcal{D}}}

\newcommand{\Lstar}{\ensuremath{L_{\ast}}}
\newcommand{\Tstar}{\ensuremath{T_{\ast}}}
\newcommand{\Rpole}{\ensuremath{R_{\rm p}}}
\newcommand{\Req}{\ensuremath{R_{\rm eq}}}
\newcommand{\Mstar}{\ensuremath{M_{\ast}}}
\newcommand{\Fstar}{\ensuremath{F_{\ast}}}

\newcommand{\area}{\ensuremath{\Sigma_{0}}}

\newcommand{\rstar}{\ensuremath{r_{\ast}}}
\newcommand{\Rstar}{\ensuremath{R_{\ast}}}

\newcommand{\Prot}{\ensuremath{P_{\rm rot}}}

\newcommand{\Lsun}{\ensuremath{{\rm L}_{\sun}}}
\newcommand{\Rsun}{\ensuremath{{\rm R}_{\sun}}}
\newcommand{\Msun}{\ensuremath{{\rm M}_{\sun}}}

\newcommand{\cmsg}{\ensuremath{{\rm cm^{2}\,g^{-1}}}}

\newcommand{\gcmc}{\ensuremath{{\rm g\,cm^{-3}}}}
\newcommand{\cms}{\ensuremath{{\rm cm^{-2}}}}
\newcommand{\gcmss}{\ensuremath{{\rm g\,cm^{-2}\,s^{-1}}}}
\newcommand{\kms}{\ensuremath{{\rm km\,s^{-1}}}}
\newcommand{\yr}{\ensuremath{{\rm yr}}}

\newcommand{\vhone}{\textsc{vh-1}}

\newcommand{\sorie}{$\sigma$~Ori~E}
\newcommand{\toric}{$\theta^{1}$~Ori~C}

\newcommand{\chandra}{\textit{Chandra}}
\newcommand{\xmm}{\textit{XMM-Newton}}
\newcommand{\iue}{\textit{IUE}}

\newcommand{\hess}{HESS}
\newcommand{\veritas}{VERITAS}

\newcommand{\apec}{\textsc{apec}}

\newcommand{\Halpha}{H$\alpha$}

\title[A Rigid-Field Hydrodynamics approach]
      {A Rigid-Field Hydrodynamics approach to modeling the magnetospheres of massive stars}
\author[R. H. D. Townsend et al.]
       {R. H. D. Townsend\thanks{E-mail: rhdt@bartol.udel.edu}, S. P. Owocki and A. ud-Doula\\
Bartol Research Institute,
Department of Physics \& Astronomy,
University of Delaware,
Newark, DE 19716, USA}

\date{%
Accepted 2007 August 27. Received 2007 August 7; in original form 2007 June 25
}

\pagerange{\pageref{firstpage}--\pageref{lastpage}}
\pubyear{2007}

\begin{document}


\maketitle

\label{firstpage}

\begin{abstract}
We introduce a new Rigid-Field Hydrodynamics approach to modeling the
magnetospheres of massive stars in the limit of very-strong magnetic
fields. Treating the field lines as effectively rigid, we develop
hydrodynamical equations describing the 1-dimensional flow along each,
subject to pressure, radiative, gravitational, and centrifugal forces. We solve
these equations numerically for a large ensemble of field lines, to
build up a 3-dimensional time-dependent simulation of a model star
with parameters similar to the archetypal Bp star \sorie. Since the
flow along each field line can be solved for independently of other
field lines, the computational cost of this approach is a fraction of
an equivalent magnetohydrodynamical treatment.

The simulations confirm many of the predictions of previous analytical
and numerical studies. Collisions between wind streams from opposing
magnetic hemispheres lead to strong shock heating. The post-shock
plasma cools initially via X-ray emission, and eventually accumulates
into a warped, rigidly rotating disk defined by the locus of minima of
the effective (gravitational plus centrifugal) potential. But a number
of novel results also emerge. For field lines extending far from the
star, the rapid area divergence enhances the radiative acceleration of
the wind, resulting in high shock velocities (up to $\sim
3,000\,\kms$) and hard X-rays. Moreover, the release of centrifugal
potential energy continues to heat the wind plasma after the shocks,
up to temperatures around twice those achieved at the shocks
themselves. Finally, in some circumstances the cool plasma in the
accumulating disk can oscillate about its equilibrium position,
possibly due to radiative cooling instabilities in the adjacent
post-shock regions.
\end{abstract}



\section{Introduction} \label{sec:intro}

During their main-sequence evolution, massive, hot stars lack the
envelope convection zones that generate magnetic fields in the Sun and
other cool stars. Nonetheless, since the 1970s it has been known that
a small, chemically peculiar subset -- the Bp stars -- possess
global-scale fields at the kilogauss level (e.g.,
\citealp{BorLan1979}; \citealp*{Bor1983}). Moreover, the significant
advances in spectropolarimetric instrumentation over the past three
decades have led to the discovery of $\sim 100-1,000\,{\rm G}$ fields
in a number of other massive stars, including two O-type stars
(\toric\ -- \citealp{Don2002}; HD~191612 -- \citealp{Don2006a}),
X-ray bright B-type stars ($\tau$~Sco -- \citealp{Don2006b}), and a
number of slowly pulsating B-type stars \citep{Hub2007}.

On the theoretical side, the genesis of massive-star fields remain the
subject of some controversy, with fossil-origin explanations
\citep[e.g.,][]{FerWik2005,FerWik2006} competing against dynamo models
involving processes such as core convection \citep{ChaMacG2001},
Tayler-Spruit instabilities \citep{MulMacD2005}, and global Rossby
modes \citep{Air2000}.  However, considerable progress has been made
in understanding how the magnetic fields channel and confine the
stars' dense, supersonic, radiatively driven winds. The seminal
Magnetically Confined Wind Shock (MCWS) model of
\citet{BabMon1997a,BabMon1997b} conjectured that wind streams from
opposing footpoints collide near the summits of closed magnetic loops,
shock-heating the plasma to temperatures $T \sim 10^{6}-10^{7}\,{\rm
K}$ at which thermal X-ray emission becomes important. Subsequent
magnetohydrodynamical (MHD) simulations by \citet{udDOwo2002}
confirmed the basic MCWS paradigm, and led to the development of a
quantitative magnetic wind-shock model for the hard X-ray emission of
\toric, which shows good agreement with \chandra\ observations of the
star \citep{Gag2005}.

MHD simulation is a powerful tool for modeling magnetic wind
confinement, but becomes increasingly difficult toward large values of
the confinement parameter $\eta$, defined as the ratio between
magnetic and kinetic energy densities \citep{udDOwo2002}. At large
$\eta$, the field lines are scarcely affected by the plasma flowing
along them. Such a rigid character implies a high Alfv\'{e}n speed,
and in turn a short numerical timestep in MHD codes to ensure Courant
stability \citep{Pre1992}. In the case of the Bp stars --
characterized by confinement parameters up to $\eta \sim 10^{7}$ --
the required timesteps are in fact too short for MHD simulation to be
practical.

For these strongly magnetic stars \citet[hereafter TO05]{TowOwo2005}
expanded on previous work by \citet{MicStu1974} and \citet{Nak1985},
to develop a Rigidly Rotating Magnetosphere (RRM) model based on the
simplifying ansatz that field lines are \emph{completely} rigid. The
RRM model does not consider the detailed physics of the wind streams
feeding into the collision shocks, but instead focuses on the fate of the
post-shock plasma after it has radiatively cooled back down to
photospheric temperatures. In a rotating star, this plasma has a
tendency to settle into magnetohydrostatic stratifications centred on
local minima of the effective (gravitational plus centrifugal)
potential. For an oblique dipole magnetic field, the locus of these
potential wells resembles a warped disk that corotates rigidly with
the star. When applied to the archetypal magnetic Bp star \sorie\
(HD~37479; B2Vpe), the \Halpha\ emission from the disk plasma shows
very good agreement with that seen in observations \citep*{Tow2005},
lending strong support to the model.

Building on the success of the RRM model, this paper presents a new
Rigid-Field Hydrodynamics (RFHD) approach to modeling massive-star
magnetospheres in the strong-field limit. We again assume that the
field lines behave as completely rigid, but we now explicitly consider
the time-dependent evolution of the magnetically channeled wind. This
approach not only furnishes a \emph{dynamical} picture of disk
accumulation, it also opens up the possibility of synthesizing
observables for the shock-heated wind plasma, across a broad range of
wavelengths extending from X-ray through to radio.

In the following section, we consider the 1-D hydrodynamical problem
of flow along each rigid field line, with an emphasis on the specific,
simple case of a dipole field topology. In \S\ref{sec:code}, we
introduce a numerical code that solves the governing equations
along many field lines, to build up a 3-D simulation of a massive-star
magnetosphere. We use this code in \S\ref{sec:calcs} to model a star
loosely based on \sorie; results from these simulations are presented
and analyzed in \S\ref{sec:results}. In \S\ref{sec:discuss}, we
examine some of the broader issues pertaining to the RFHD approach,
and in \S\ref{sec:summary} we summarize the paper.


\section{Rigid field hydrodynamics} \label{sec:rfhd}

As we discuss above, the key notion of the RFHD approach is that the
magnetic field at sufficiently high strengths behaves as if it were
rigid. This rigid field is anchored to the star, and co-rotates with
it. Under the frozen flux condition of ideal MHD, plasma is
constrained to flow along field lines, and it therefore describes
trajectories that are \emph{fixed} in the co-rotating frame.

The shape of these trajectories is specified \emph{a priori} by the
chosen magnetic topology. However, the plasma state (density,
temperature, velocity, etc.) along each field line is determined by
the 1-D hydrodynamical problem of flow along a tube with changing
cross-sectional area. Here, the `tube' can be identified explicitly
with a magnetic flux tube, whose area varies inversely with the local
magnetic flux density $B \equiv |\bB|$ in order to ensure that $\nabla
\cdot \bB = 0$. The character of the flow is dictated primarily by the
forces that act to accelerate or decelerate the plasma: pressure
gradients, gravity, the centrifugal force, and radiative
forces. Perhaps surprisingly, magnetic and Coriolis forces play no
direct role in the 1-D flow problem, because they are always directed
perpendicular to the instantaneous velocity vector \bv. (This vector
is itself everywhere parallel to the field-line tangent vector $\bes
\equiv \bB/B$.) In fact, these forces act similarly to the centripetal
force of a circular orbit, furnishing a net acceleration perpendicular
to \bv\ that leads to curved plasma trajectories yet does no work.

\subsection{Euler equations} \label{ssec:rfhd-euler}

To elaborate on the foregoing discussion, we introduce the Euler
equations in conservation form for the 1-D hydrodynamical problem
comprising RFHD,
\begin{gather} 
\label{eqn:hydro-mass}
\frac{\partial \rho}{\partial t} + \frac{1}{A}
\frac{\partial}{\partial s} \left(A \rho v\right) = 0, \\
\label{eqn:hydro-mom}
\frac{\partial \rho v}{\partial t} + \frac{1}{A}
\frac{\partial}{\partial s} \left(A \rho v^{2}\right) + \frac{\partial
  p}{\partial s} = \rho (\bgeff + \bgrad) \cdot \bes, \\
\label{eqn:hydro-energy}
\frac{\partial \rho e}{\partial t} + \frac{1}{A}
\frac{\partial}{\partial s} \left[A v \left(\rho e + p\right)\right] = \rho v
(\bgeff + \bgrad) \cdot \bes + \cool.
\end{gather}
Here, the independent variables are the arc distance $s$ along the
field line (relative to some arbitrary zero point) and time $t$, while
the dependent variables are density $\rho$, pressure $p$, velocity $v
\equiv |\bv|$ and total energy per unit mass $e$. The term $A(s)$
describes the spatially varying cross-sectional area of the flow tube,
and depends on the magnetic topology; expressions for this term in the
case of a dipole field are derived in the following section. The
acceleration vectors \bgeff\ and \bgrad\ are due to the effective
gravity and the radiative line force, respectively, and are considered
in greater detail in~\S\ref{ssec:rfhd-pot}
and~\S\ref{ssec:rfhd-rad}. Inter-relationships between $\rho$, $p$,
$v$ and $e$ are determined from equations of state and total energy,
defined in \S\ref{ssec:rfhd-state}. Finally, the term \cool\ is the
volumetric energy loss rate due to cooling processes, and is discussed
in \S\ref{ssec:rfhd-cool}. For reasons elaborated there, we do
not include the effects of thermal conduction in the energy
conservation equation~(\ref{eqn:hydro-energy}).

By setting all velocities and time derivatives to zero, the momentum
conservation equation~(\ref{eqn:hydro-mom}) reduces to the condition
of magnetohydrostatic equilibrium, in which body forces are balanced
by pressure gradients. Because it furnishes the basis of the precursor
RRM model, we review this static limit in Appendix~\ref{app:static}.

\subsection{Dipole field geometry} \label{ssec:rfhd-geom}

\begin{figure}
\begin{center}
\includegraphics{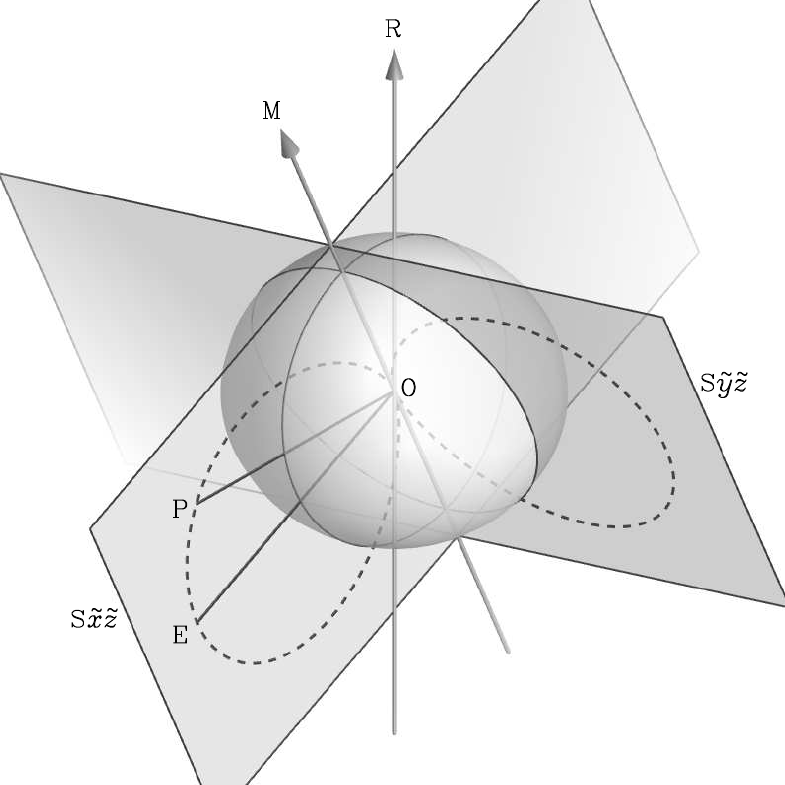}
\end{center}
\caption{An illustration of the oblique dipole geometry described in
the text. The star is shown as an oblate spheroid centred on the
origin O, with magnetic axis M and rotational axis R; the angle MOR is
the obliquity $\beta$. The surfaces S$\xmag\zmag$ and S$\ymag\zmag$
are the \xmag--\zmag\ and \ymag--\zmag\ planes of the magnetic
reference frame, corresponding to azimuths $\pmag=(0\degr,180\degr)$
and $\pmag=(90\degr,270\degr)$, respectively. Two selected field
lines, having $\pmag=0\degr$ and $\pmag=90\degr$, are shown by dotted
lines; on the $\pmag=0\degr$ field line, E labels the magnetic
equator, and P a point on the field line. The line OE (measured by
convention in units of the rotational polar radius \Rpole) gives the
magnetic shell parameter $L$; likewise, EP defines the arc distance
coordinate $s$ of the point P. (In this case, $s<0$, because P lies in
the northern magnetic hemisphere). OP is the radial coordinate
$\rmag=\rrot$ of P, and the angle MOP (ROP) gives the corresponding
colatitude \tmag\ (\trot) in the magnetic (rotational) reference
frame.}
\label{fig:geometry}
\end{figure}

Although the RFHD approach is in principle applicable to
arbitrary magnetic topologies, the present study focuses on the
simple case of an oblique dipole field. Let $(\rrot,\trot,\prot)$ be
the spherical polar coordinates in the reference frame aligned with
the rotation axis; likewise, let $(\rmag,\tmag,\pmag)$ be the
corresponding coordinates in the frame aligned with the dipole
magnetic axis. (This is the same notation as adopted in TO05). As
illustrated in Fig.~\ref{fig:geometry}, the magnetic axis is tilted
with respect to the rotation axis by the magnetic obliquity $\beta$.

In the magnetic reference frame, the magnetic flux vector is
expressed as
\begin{equation} \label{eqn:B}
\bB = \frac{\Bz}{2 (\rmag/\Rpole)^{3}} \left( 2 \cos\tmag \,\bermag +
\sin\tmag \,\betmag \right)\, ,
\end{equation}
where \Bz\ sets the overall field strength\footnote{For an aligned
dipole ($\beta=0$), \Bz\ corresponds to the polar field strength;
however, this does not generally hold when $\beta>0$, due to the
oblateness of the star (\S\ref{ssec:rfhd-star}).}, \Rpole\ is the
rotational polar radius (a convenient normalizing length), and
\bermag\ and \betmag\ are the unit basis vectors in the magnetic
radial and polar directions, respectively. From this expression, the
tangent vector \bes\ is obtained as
\begin{equation}
\bes = \frac{\bB}{B} = \frac{1}{\sqrt{1 + 3\cos^{2}\tmag}} \left( 2\cos\tmag \,\bermag
+ \sin\tmag \,\betmag \right)\, .
\end{equation}
This result also follows from the parametric equation for a dipole
field line,
\begin{equation} \label{eqn:field-r}
\frac{\rmag}{\Rpole} = L \sin^{2}\tmag
\end{equation}
\citep[e.g.,][]{Nak1985,BabMon1997a}, where the magnetic shell
parameter $L$ measures the maximal radius reached by the field line,
in units of \Rpole. To label a field line uniquely, it suffices to
specify $L$ and the magnetic azimuthal coordinate \pmag\ defining the
half-plane that contains the line.

The spatial variable $s$ in the 1-D hydrodynamical
equations~(\ref{eqn:hydro-mass}--\ref{eqn:hydro-energy}) is the arc
distance along each field line, and is found from the dipole line
element
\begin{equation} \label{eqn:line-element}
\diff s^{2} = \diff \rmag^{2} + \rmag^{2}\,\diff\tmag^{2}
= L^{2} \Rpole^{2} \sin^{2}\tmag \left(1 + 3\cos^{2}\tmag\right)
\,\diff\tmag^{2}\, .
\end{equation}
Solving this differential equation for $s$, we obtain
\begin{multline} \label{eqn:s}
\frac{s}{\Rpole} = -\frac{L}{2} \left[
  \frac{\sinh^{-1}\left(\sqrt{3}\cos\tmag\right)}{\sqrt{3}} + \mbox{}
  \right. \\
\left. \cos\tmag \sqrt{1 + 3\cos^{2}\tmag}
\right]\, ,
\end{multline}
where the constant of integration is chosen to place the origin $s=0$
at the magnetic equator, $\tmag=90\degr$.  The negative sign on the
right-hand side arises from selecting the positive root of
eqn.~(\ref{eqn:line-element}), so that $s$ increases in the same
direction as \tmag, and $s$ is negative (positive) in the northern
(southern) magnetic hemisphere. Since this equation is transcendental,
calculation of the inverse function $\tmag(s)$ must be undertaken
numerically (\S\ref{ssec:code-coord}).

The footpoints of the field line in the northern and southern magnetic
hemispheres are denoted as \sstarN\ and \sstarS, respectively. For a
spherical star with radius \Rstar,
\begin{multline} \label{eqn:sstar-spherical}
\frac{\sstarN}{\Rstar} = - \frac{\sstarS}{\Rstar} = - \frac{L}{2} \left[
  \frac{\sinh^{-1}\sqrt{3 - 3/L}}{\sqrt{3}} + \mbox{} \right. \\
\left. \sqrt{1 - 1/L}\sqrt{4 - 3/L}
\right]\, .
\end{multline}
However, for an oblate star (cf.~\S\ref{ssec:rfhd-star}) calculation
of \sstarN\ and \sstarS\ once again must proceed numerically
(\S\ref{ssec:code-coord}).

The variation in the cross-sectional area of each flow tube is
determined by the requirement of magnetic flux conservation. If $\Aeq$
is the area at the magnetic equator, then the area $A$ at any
colatitude \tmag\ must satisfy
\begin{equation}
A B = \Aeq\,\Beq\, ,
\end{equation}
where
\begin{equation}
\Beq = \frac{\Bz}{2 L^{3}}
\end{equation}
is the field strength at the magnetic equator. Eliminating $B$ with
the help of eqns.~(\ref{eqn:B}) and~(\ref{eqn:field-r}), we find that
\begin{equation} \label{eqn:A}
\frac{A}{\Aeq} = \frac{\sin^{6}\tmag}{\sqrt{1 + 3\cos^{2}\tmag}}\, .
\end{equation}
Combined with the inverse function $\tmag(s)$, this expression is
used to construct the area function $A(s)$ appearing in the Euler
equations~(\ref{eqn:hydro-mass}--\ref{eqn:hydro-energy}).

In addition to the area function $A$, finite-volume hydrodynamical
codes such as \vhone\ (cf.~\S\ref{sec:code}) require specification of
the volume function
\begin{equation}
V \equiv \int A \,\diff s\, .
\end{equation}
To obtain $V$ in the present case, we note from
eqns.~(\ref{eqn:line-element}) and~(\ref{eqn:A}) that
\begin{equation}
A \,\diff s = A \frac{\diff s}{\diff \tmag} \,\diff\tmag
= L \Aeq \Rpole \sin^{7}\tmag \,\diff\tmag\, .
\end{equation}
It therefore follows that
\begin{multline} \label{eqn:V}
\frac{V}{\Aeq \Rpole} = L \left[ -\frac{7}{320} \left(
25 \cos\tmag - 5 \cos3\tmag + \cos5\tmag
\right) + \mbox{} \right. \\
\left. \frac{1}{448} \cos7\tmag \right]\, ,
\end{multline}
and, as with $A(s)$, the inverse function $\tmag(s)$ is used to find
$V(s)$.

\subsection{Effective potential} \label{ssec:rfhd-pot}

The effective gravitational acceleration \bgeff\ in
eqns.~(\ref{eqn:hydro-mom},\ref{eqn:hydro-energy}) combines the
Newtonian gravity with the centrifugal force associated with enforced
co-rotation. Together, these forces are derived from a scalar
effective potential \poteff,
\begin{equation} \label{eqn:geff}
\bgeff = -\nabla \poteff\, .
\end{equation}
Within the Roche (point-mass) approximation, this effective potential
is given by
\begin{equation} \label{eqn:roche}
\poteff = - \frac{G \Mstar}{r} - \frac{1}{2} \Omega^{2} \drot^{2}\, .
\end{equation}
Here, \Mstar\ is the stellar mass, $\Omega$ the angular rotation
frequency, and $\drot \equiv \rrot \sin \trot$ is the distance from
the rotation axis. Combining the above two expressions, the effective
gravity term in the momentum and energy conservation
equations~(\ref{eqn:hydro-mom},\ref{eqn:hydro-energy}) is found as
\begin{equation}
\bgeff \cdot \bes = -\frac{G \Mstar}{r^{2}} \,\berrot\cdot\bes +
\Omega^{2} \drot \,\bedrot\cdot\bes\, ,
\end{equation}
where \berrot\ and \bedrot\ are the unit vectors in the \rrot\ and
\drot\ directions, respectively. For a dipole field
(\S\ref{ssec:rfhd-geom}),
\begin{equation} \label{eqn:chi}
\berrot \cdot \bes = \bermag \cdot \bes = \frac{2 \cos\tmag}{\sqrt{1 +
    3\cos^{2}\tmag}} \equiv \cos \chi 
\end{equation}
gives the component of the local radial vector projected along the
field line. The angle $\chi$ introduced in this expression, being that
between the field line and the radial vector, also appears below in
the equations governing the radiative acceleration.

\subsection{Stellar properties} \label{ssec:rfhd-star}

In addition to generating the effective gravity, \poteff\ determines
the shape and surface properties of the oblate, centrifugally
distorted star. In the Roche approximation (eqn.~\ref{eqn:roche}), the
surface is an equipotential whose radius \rstar\ varies with
rotational colatitude \trot\ as
\begin{equation} \label{eqn:rstar}
\frac{\rstar}{\Rpole} = \frac{3}{w \sin\trot} \cos \left[\frac{\pi +
    \cos^{-1}(w \sin\trot)}{3}\right]
\end{equation}
\citep[e.g.,][]{Cra1996}. Here,
\begin{equation} \label{eqn:rot-freq}
w \equiv \Omega\,\sqrt{\frac{27 \Rpole^{3}}{8 G \Mstar}}
\end{equation}
is the normalized rotation angular frequency, with $w = 1$
corresponding to critical rotation.

Due to gravity darkening \citep{vonZ1924}, the flux emitted by a
rotating, radiative stellar envelope varies in proportion to the local
effective gravity $|\bgeff|$. However, in evaluating the radiative
acceleration (\S\ref{ssec:rfhd-rad}) and the rate of inverse Compton
cooling (\S\ref{ssec:rfhd-state}), we choose for simplicity to treat
the circumstellar radiation field as originating from a
spherically-symmetric point source of luminosity \Lstar. For
consistency, we therefore neglect the variation of the surface flux,
setting it to the constant value
\begin{equation} \label{eqn:flux}
\Fstar \equiv \frac{\Lstar}{\area}\, .
\end{equation}
Here, \area\ is the total surface area of the oblate star, which can
be approximated to within 2\% by the polynomial
\begin{multline}
\area = 4 \pi \Rpole^{2} \left( 1 + 0.19444 w^{2} + 0.28053 w^{4} - 1.9014
w^{6} + \mbox{} \right. \\
\left. 6.8298 w^{8} - 9.5002 w^{10} + 4.6631 w^{12} \right)
\end{multline}
\citep{Cra1996}. From \Fstar\ and the Stefan-Boltzmann law, a nominal
stellar surface temperature is defined as
\begin{equation} \label{eqn:Tstar}
\Tstar = \left(\frac{\Fstar}{\sigma}\right)^{1/4}\, .
\end{equation}

\subsection{Radiative acceleration} \label{ssec:rfhd-rad}

To obtain an expression for the radiative acceleration \bgrad\, we
employ the \citet*[hereafter CAK]{Cas1975} formalism for line-driven
stellar winds. For simplicity, the star is treated as a point source
of radiation, giving an acceleration
\begin{equation} \label{eqn:bgrad}
\bgrad = \frac{1}{1 - \alpha} \frac{\kappae \Lstar \Qbar}{4\pi r^{2}
  c} \left(\frac{|\deltav|}{\rho c \Qbar \kappae}\right)^{\alpha} \,\berrot
\end{equation}
\citep[e.g.,][]{Owo2004}. Here, $\alpha$ is the CAK power-law index;
\Qbar\ is the dimensionless line strength parameter introduced by
\citet{Gay1995}; \Lstar\ is the stellar luminosity
(\S\ref{ssec:rfhd-star}); and \kappae\ is the electron-scattering
opacity, $\approx 0.34\,\cmsg$ for a fully ionized solar-composition
plasma. The term \deltav, a measure of the local velocity gradient
that determines the \citet{Sob1960} optical depth, is defined as
\begin{equation}
\deltav \equiv \berrot \cdot \nabla (\berrot \cdot \bv)\, .
\end{equation}
From eqn.~(\ref{eqn:chi}),
\begin{equation}
\berrot \cdot \bv = v \cos\chi\, ,
\end{equation}
and thus
\begin{equation}
\deltav = \berrot \cdot \nabla (v \cos\chi) = \cos\chi \,\berrot \cdot
\nabla v\, .
\end{equation}
(The second equality follows because $\chi$ is a function of \tmag\
alone, and therefore commutes with the radial directional derivative
$\berrot \cdot \nabla$). Expanding out the gradient operator in the
magnetic reference frame, we obtain
\begin{equation} \label{eqn:deltav-full}
\deltav = \cos\chi \left(\frac{\partial v}{\partial
  \rmag}\right)_{\tmag}\, .
\end{equation}

To evaluate the radial derivative in this latter equation, 
it is necessary to know the flow velocity on field lines adjacent to
the one under consideration. 
In order to avoid this complication, we make the simplifying 
assumption that the polar velocity derivate vanishes, 
\begin{equation} \label{eqn:dvdtheta-approx}
\left(\frac{\partial v}{\partial \tmag}\right)_{\rmag} \approx 0\, .
\end{equation}
As we demonstrate in Appendix \ref{app:velocity}, this assumption
implies that
\begin{equation} \label{eqn:dvdr-approx}
\left(\frac{\partial v}{\partial \rmag}\right)_{\tmag}  =
\left(\frac{\partial v}{\partial s}\right)_{L} \sec\chi \, .
\end{equation}
Substituting this expression back into eqn.~(\ref{eqn:deltav-full}) yields
\begin{equation} \label{eqn:deltav-approx}
\deltav = \left(\frac{\partial v}{\partial s}\right)_{L}\, ,
\end{equation}
which in combination with eqn.~(\ref{eqn:bgrad}) leads to the
final expression for the radiative acceleration,
\begin{equation} \label{eqn:bgrad-approx}
\bgrad = \frac{1}{1 - \alpha} \frac{\kappae \Lstar \Qbar}{4\pi r^{2}
  c} \left(\frac{|\partial v/\partial s|}{\rho c \Qbar
  \kappae}\right)^{\alpha} \,\berrot\, .
\end{equation}
To conform with our earlier notation
(cf.~\S\ref{ssec:rfhd-euler}), here we have dropped the `${L}$'
subscript on the velocity gradient $\partial v/\partial s$.

The central approximation here,
eqn. (\ref{eqn:dvdtheta-approx}), is equivalent to assuming that $v$
-- which might more correctly be termed the flow \emph{speed} --
depends only on \rmag, although of course the local flow
\emph{direction} still depends on \tmag\ since it follows the
field-line orientation.  Analysis based on 2-D MHD simulations
\citep[cf.][]{OwoudD2004} indicates that this approximation is
well-justified in the regions of rapid wind acceleration near to the
star.  Moreover, in the present work we have carried out a posteriori
checks using an ensemble of neighboring 1-D RFHD calculations, and
find that the error in \bgrad\ arising from using
eqn.~(\ref{eqn:deltav-approx}) for \deltav\ instead of the exact
expression~(\ref{eqn:deltav-full}) is typically 10\% or less
throughout the radiatively driven regions of the magnetosphere.

We emphasize that the benefit from this approximation is
significant.  Because eqn.~(\ref{eqn:bgrad-approx}) depends only on
the velocity gradient \emph{along} a field line, it allows the flow to
be modeled completely \emph{independently} of other field lines. Not
only is this advantageous in terms of computational efficiency, it
also simplifies the interpretation and analysis of simulation results
(\S\ref{sec:results}).

\subsection{Equations of state and energy} \label{ssec:rfhd-state}

We assume an ideal gas, so that
\begin{equation} \label{eqn:state}
p = \frac{\rho k T}{\mu \matom},
\end{equation}
with $T$ the temperature and \matom\ the atomic mass unit. The
mean molecular weight $\mu$ is determined from an expression
appropriate to a fully ionized mixture,
\begin{equation}
\mu = \left[ 2 X + \frac{3}{4}(1 - X - Z) + \frac{Z}{2} \right]^{-1},
\end{equation}
where $X$ and $Z$ are the usual hydrogen and metal mass fractions. The
accompanying equation for the specific (per-unit-mass) total energy is
\begin{equation} \label{eqn:energy}
e = \frac{1}{\gamma-1}\frac{k T}{\mu \matom} + \frac{v^{2}}{2},
\end{equation}
where $\gamma$ is the ratio of specific heats.

\subsection{Cooling and thermal conduction}\label{ssec:rfhd-cool}

The volumetric cooling rate \cool\ in the energy conservation
equation~(\ref{eqn:hydro-energy}) is evaluated as the sum of two
terms,
\begin{equation} \label{eqn:coolfunc}
\cool = \coolat + \coolic\, ,
\end{equation}
representing contributions from atomic processes and inverse Compton
scattering by thermal electrons, respectively. The first term
is calculated from
\begin{equation} \label{eqn:coolat}
\coolat = \nel \np \coolcoeff(T)\, ,
\end{equation}
where
\begin{equation} \label{eqn:np}
\np = \frac{X}{\matom} \rho
\end{equation}
and
\begin{equation} \label{eqn:nel}
\nel = \frac{1+X}{2 \matom} \rho
\end{equation}
define the proton and electron number densities, respectively. The
temperature-dependent cooling coefficient \coolcoeff\ is obtained from
the curve published by \citet{MacDBai1981}.

The inverse Compton term \coolic\ is evaluated with the aid of
eqn.~(4) of \citet{WhiChe1995},
\begin{equation}
\coolic = 4 \frac{\kappae}{c} \nel \Uph k T\, .
\end{equation}
(Note that in their eqn.~5, these authors define the symbol \coolic\
differently than above). Here,
\begin{equation}
\Uph = \frac{\Lstar}{4 \pi r^{2} c}
\end{equation}
is the energy density associated with the star's radiation field,
evaluated in the same point-star limit as the radiative acceleration
(\S\ref{ssec:rfhd-rad}).

We have elected to neglect the effects of thermal conduction
in the energy conservation equation, because to include these properly
requires addressing a large number of issues that are beyond the
present scope.  These center on uncertainties both in the treatment of
departures from the classical \citet{Spi1962} thermal conductivity
\citep[e.g.,][]{LevEic1992, PisEic1998}, and in the incorporation of
heat flux saturation near shocks and other regions of steep
temperature gradients \cite[see][and references therein]{Lac1988,
BanChe1994a}.  Moreover, apart from these physical difficulties,
proper inclusion of conduction in a hydrodynamical code is a
challenging task, particularly in the context of accurately modeling
the kind of strong shocks that occur in our simulations \citep[see,
e.g.,][]{Rea1995}. We thus defer consideration of thermal conduction
to future work.



\section{The RFHD Code} \label{sec:code}

As discussed in \S\ref{ssec:rfhd-rad}, the utility of the approximate
expression~(\ref{eqn:deltav-approx}) for the velocity gradient term
\deltav\ lies in the fact that the flow along each individual field
line may be simulated completely independently of other field
lines. Thus, by performing many separate 1-D simulations for differing
field lines and piecing them together, a 3-D hydrodynamical
\emph{meta}simulation of a massive-star magnetosphere can be built up
at a fraction of the computational cost of an equivalent MHD
calculation.

Each individual 1-D simulation requires solution of the Euler
equations~(\ref{eqn:hydro-mass}--\ref{eqn:hydro-energy}), and for this
we employ a customized version of the \vhone\ hydrodynamical code
developed by J. Blondin and colleagues. \vhone\ is a finite volume
code based on the Lagrangian version of the piecewise parabolic method
(PPM) devised by \citet{ColWoo1984}. The modifications to \vhone\
primarily encompass incorporation of the dipole geometry
(\S\ref{ssec:rfhd-geom}), acceleration
terms~(\S\S\ref{ssec:rfhd-pot},\ref{ssec:rfhd-rad}) and cooling
(\S\ref{ssec:rfhd-cool}) specific to the RFHD problem.  We do not
discuss these modifications in detail, but in the following sections
we highlight specific issues that arose during the code development
phase.

\subsection{Grid design} \label{ssec:code-grid}

The Eulerian grid in \vhone\ must be designed with care, to ensure
that regions of physical interest are properly resolved, and to avoid
the generation of spurious numerical instabilities. To this end, we
divide the grid into three domains. The two `surface' domains are each
composed of a fixed number \Nsurf\ of zones, extending from the
surface footpoints $s=\sstarN,\sstarS$ (cf.~\S\ref{ssec:code-coord})
to somewhat above the expected position of the sonic point $|v| =
\csoundstar$. (Here $\csoundstar \equiv \sqrt{k \Tstar/\mu \matom}$
denotes the isothermal sound speed at the surface, with \Tstar\ the
stellar surface temperature introduced in \S\ref{ssec:rfhd-star}.)
These zones are non-uniform, with the size of each zone being 1.1
times that of its neighbour closer to the stellar surface. With their
high spatial resolution, the surface domains are designed to resolve
the smooth transition of the wind from a subsonic, near-hydrostatic
state to a supersonic outflow.

The third, `magnetosphere' domain spans the physically interesting
regions of the circumstellar environment, and is composed of \Nmag\
zones of uniform size $\diff s \approx (\sstarS-\sstarN)/\Nsurf$, that
bridge between the two thin surface domains. In determining an
appropriate value for \Nmag, we are motivated by a desire to resolve
the dense, co-rotating disk predicted by the RRM model to accumulate
at local minima of the effective potential (see TO05; see also
Appendix~\ref{app:static}). As demonstrated in the Appendix, the scale
height of this disk is independent of the magnetic shell parameter $L$
when $L$ becomes large. Thus, we vary \Nmag\ with $L$ according to the
simple formula
\begin{equation}
\Nmag = \operatorname{int}(\deldisk L + \delstar L^{-1/2})\, ,
\end{equation}
where $\operatorname{int}()$ denotes the integer part. The first term in the
parentheses ensures that the disk remains properly resolved at large
$L$, with a constant number of zones per asymptotic scale height
\hinf\ (cf. eqn.~\ref{eqn:hinf}); while the second is an ad-hoc
addition that increases the spatial resolution near the star. The
parameters \deldisk\ and \delstar\ are adjusted to achieve the desired
spatial resolution in the far- and near-star limits.

\subsection{Field-line coordinates} \label{ssec:code-coord}

To determine the arc distance coordinates $(\sstarN,\sstarS)$ of the
field-line footpoints, required in setting up the Eulerian grid
(\S\ref{ssec:code-grid}), we solve the equation
\begin{equation}
\rstar = L \Rpole \sin^{2} \tmag
\end{equation}
to find the magnetic colatitudes $(\tmagN,\tmagS)$ associated with
each footpoint. (The footpoints are not mirror symmetric about the
magnetic equator unless $w = 0$ or $\beta = 0\degr$). The
corresponding values of $s$ then follow from eqn.~(\ref{eqn:s}). In
the above equation, \rstar\ is a function of the rotational colatitude
\trot\ (cf. eqn.~\ref{eqn:rstar}); thus, it depends implicitly on
\tmag\ and \pmag\ via the coordinate transformation between magnetic
and rotational reference frames (see Fig.~\ref{fig:geometry}). To
solve this equation, we use Brent's algorithm \citep{Pre1992}.

Once the Eulerian grid is established, we calculate \tmag\ at each zone
boundary from the corresponding $s$ values, by
inverting eqn.~(\ref{eqn:s}). For this, we use a Newton-Raphson
iteration operating on $\cos\tmag$, starting from the initial guess
\begin{equation}
\cos\tmag = -s \left[ L \Rpole \left(\frac{\sinh^{-1}\sqrt{3}}{2
    \sqrt{3}} + 1\right) \right]^{-1}\, .
\end{equation}
After each Lagrangian step of the PPM algorithm, the advection of
zone boundaries means that the \tmag\ values must be updated. The
Newton-Raphson iteration is in this case too slow to be useful, so we
instead evaluate \tmag\ by cubic-spline interpolation \citep{Pre1992}
from the stored Eulerian-grid values.

\subsection{Initial state} \label{ssec:code-initial}

The initial, $t=0$ state for simulations is based on a spherically
symmetric, accelerating wind. The velocity is obtained by projecting a
canonical velocity law onto field lines:
\begin{equation}
v = \vinf \left[ 1 - \frac{\rstar}{r} \right]^{1/2} \,\cos\chi\, .
\end{equation}
Here, \vinf\ is the terminal velocity obtained from a non-rotating,
point-star CAK model,
\begin{equation} \label{eqn:vinf}
\vinf = \sqrt{\frac{\alpha}{1-\alpha} \frac{2 G \Mstar}{\Rpole}}\, .
\end{equation}
The density is likewise obtained from the condition of steady
spherical outflow,
\begin{equation}
\rho = \frac{\Mdot}{4\pi r^{2} v}\, ,
\end{equation}
where now
\begin{equation} \label{eqn:mdot}
\Mdot = \frac{\Lstar}{c^{2}} \frac{\alpha}{\alpha-1}
\left[\frac{\Qbar \Gammae}{1 - \Gammae}\right]^{(1-\alpha)/\alpha}\, ,
\end{equation}
is the CAK mass-loss rate, with
\begin{equation}
\Gammae \equiv \frac{\kappae \Lstar}{4\pi G \Mstar c}
\end{equation}
the Eddington parameter associated with electron scattering. By
assuming an isothermal initial flow, with $T = \Tstar$, the pressure
$p$ is obtained from the equation of state~(\ref{eqn:state}).

\subsection{Boundary conditions} \label{ssec:code-bound}

Boundary conditions are implemented in \vhone\ by setting dependent
variables ($\rho$, $p$, $v$, $e$) in ghost zones at both ends of the
grid. To allow for the possibility of outflow \emph{and} inflow, we
fix only the density and pressure in the ghost zones, as $\rho =
\rhostar$ and $p = \pstar$. The stellar density \rhostar\ can be
chosen with some degree of latitude, so long as it remains appreciably
above the wind density
\begin{equation}
\rhosonic = \frac{\Mdot}{4\pi\rstar^{2} \csoundstar}
\end{equation}
at the sonic point; we find that a choice $\rhostar = 10\,\rhosonic$
gives a smooth and steady wind outflow. The corresponding stellar
pressure \pstar\ is obtained from the equation of
state~(\ref{eqn:state}), with $\rho = \rhostar$ and $T = \Tstar$. The
velocity in the ghost zones is linearly extrapolated from the first
pair of computational zones (subject to the constraint that $|v|$ does
not exceed \csoundstar), and the total energy per unit mass is
evaluated using eqn.~(\ref{eqn:energy}).

\subsection{Cooling} \label{ssec:code-cool}

Cooling in the customized \vhone\ is time-split from the rest of the
PPM algorithm. Across each timestep $\diff t$, a new temperature
$T_{j}^{\dagger}$ in the $j$'th zone is calculated from the current
temperature $T_{j}$ and density $\rho_{j}$ via
\begin{equation}
T_{j}^{\rm \dagger} = T_{j} + \frac{(\gamma-1) \mu \matom}{k} \frac{
  \cool(T_{j}^{\dagger},\rho_{j}) + \cool(T_{j},\rho_{j})}{2 \rho_{j}}
  \,\diff t\, ,
\end{equation}
where $\Lambda(T,\rho)$ is the cooling rate discussed in
\S\ref{ssec:rfhd-cool}. Since this equation is implicit, the code uses
3-step iteration to converge toward an approximate value for
$T_{j}^{\dagger}$. In zones where the temperature would drop below the
stellar surface temperature \Tstar, it is reset to \Tstar; this
reflects the tendency for the photospheric radiation field to keep
circumstellar plasma warm. The resulting $T_{j}^{\dagger}$ values are
used to update the pressure in each zone, through the equation of
state~(\ref{eqn:state}).

\section{Calculations} \label{sec:calcs}

\begin{table}
\begin{tabular}{ccccccc}
\Mstar  & \Rpole  & \Tstar & \Lstar     & $\beta$    & $X$ & $Z$  \\
(\Msun) & (\Rsun) & (K)    & (\Lsun)    & (\degr)    &     &      \\ \hline
8.9     & 5.3     & 22,500 & 7,420      & 55         & 0.7 & 0.02 \\
\\
\\
\Prot   & \Qbar   & $\alpha$ & $\gamma$ & \Nsurf & \deldisk & \delstar \\
(days)  &         &          &          &        &          &          \\ \hline
1.2     & 500     & 0.6      & 5/3      & 25     & 100      & 100
\end{tabular}
\caption{Parameters for the RFHD simulation described in
\S\ref{sec:calcs}.} \label{tab:params}
\end{table}

As an initial test of the code discussed in the preceding sections,
and to explore the RFHD approach in general, we develop a model for
the magnetosphere of an oblique-dipole star whose parameters
(Table~\ref{tab:params}) loosely coincide with those of \sorie. (We
stress that we are not attempting to fine-tune a model for \sorie;
however, it makes sense to begin our \emph{qualitative} investigations
in the same approximate region of parameter space as this archetypal
star.) The mass, radius and surface temperature of the star are taken
from the recent study by \citet{Krt2006}, with the luminosity
calculated from eqns.~(\ref{eqn:flux}) and~(\ref{eqn:Tstar}). The wind
parameters \Qbar\ and $\alpha$ are assigned values typical to hot
stars \citep[e.g.,][]{Gay1995}, resulting in a CAK mass loss rate
(eqn.~\ref{eqn:mdot}) of $3.7 \times 10^{-9} \,\Msun\,\yr^{-1}$. The
rotation period \Prot\ is based on the photometric period measured by
\citet*{Hes1977}, and leads to a normalized rotation frequency $w =
0.73$. The dipole obliquity $\beta$ is adopted from the RRM model
presented by \citet{Tow2005}.

To piece together a 3-D model for the magnetosphere, we use \vhone\ to
simulate the flow along a total of 1140 field lines, arranged on a
grid of 60 magnetic shell parameters $L$ and 19 magnetic azimuths
\pmag. The azimuth values span the $0\degr \leq \pmag \leq 90\degr$
quadrant in 5\degr\ increments; symmetry is used to replicate the
simulation results over the remaining $90\degr \leq \pmag \leq
360\degr$ interval, effectively leading to 72 azimuthal points. The 60
values of the shell parameter are distributed according to the formula
\begin{equation} \label{eqn:req-grid}
\ln L = \ln 1.2  + \ln (11.2/1.2) \frac{\ell-1}{59}
\qquad (\ell = 1\ldots60)\, .
\end{equation}
This logarithmic distribution provides higher resolution close to the
star, where we expect magnetospheric structure to depend more
sensitively on $L$. The innermost, $\ell=1$ field lines have $L =
1.2$; by comparison, the stellar equatorial radius for the adopted
parameters is $\Req = 1.11\,\Rpole$. These field lines are composed of
$\Nsurf=25$ grid zones in each of the two surface domains and -- with
the choices \deldisk\ and \delstar\ listed in Table~\ref{tab:params}
-- $\Nmag=211$ zones in the intervening magnetosphere domain. The
outermost, $\ell=60$ field lines have $L = 11.2$, and therefore extend
out to just over six times the Kepler co-rotation radius $\rKep =
1.86\,\Rpole$ (cf. eqn.~\ref{eqn:rKep}). These field lines again have
$\Nsurf=25$, but now the magnetosphere domain has $\Nmag=1149$ zones,
ensuring a zone size $\diff s = 0.027\,\Rpole$ that is less than half
the asymptotic scale height $\hinf = 0.064\,\Rpole$ defined by
eqn.~(\ref{eqn:hinf}).

Each 1-D simulation runs for 200 stellar rotation cycles ($\sim
21\,{\rm Ms}$). This is significantly longer than typical flow times
$\sim 20\,{\rm ks}$, to ensure that radiative relaxation from the
initial state is obtained, and to follow the long-term evolution of
the magnetosphere. Since the simulations for each field line are
independent, they can be distributed across different processors,
computers or even clusters. In the present case, the 1140 individual
simulations were run on a cluster of eight 2.0 GHz dual-core AMD
Opteron nodes (for a total of 16 processors). On average, each
simulation took $\sim 200$ minutes (although there was a significant
spread about this mean), giving a total computation time of $\sim 10$
days.


\section{Results} \label{sec:results}

The complete 3-D magnetospheric model we describe above, composed of
1140 1-D simulations, occupies $\sim 36\,{\rm Gb}$ of
storage. Presenting this significant dataset in an informative manner
poses quite a challenge. Fortunately, the independence of the
individual simulations is once again of benefit, since the flow can be
analyzed first in one dimension, along an individual field line
(\S\ref{ssec:results-1D}); then in two dimensions, along field lines
lying in the same meridional plane (\S\ref{ssec:results-2D}); and then
in three dimensions (\S\ref{ssec:results-3D}), for the complete model.

\subsection{1-D, along individual field line} \label{ssec:results-1D}

\begin{figure*}
\begin{center}
\includegraphics{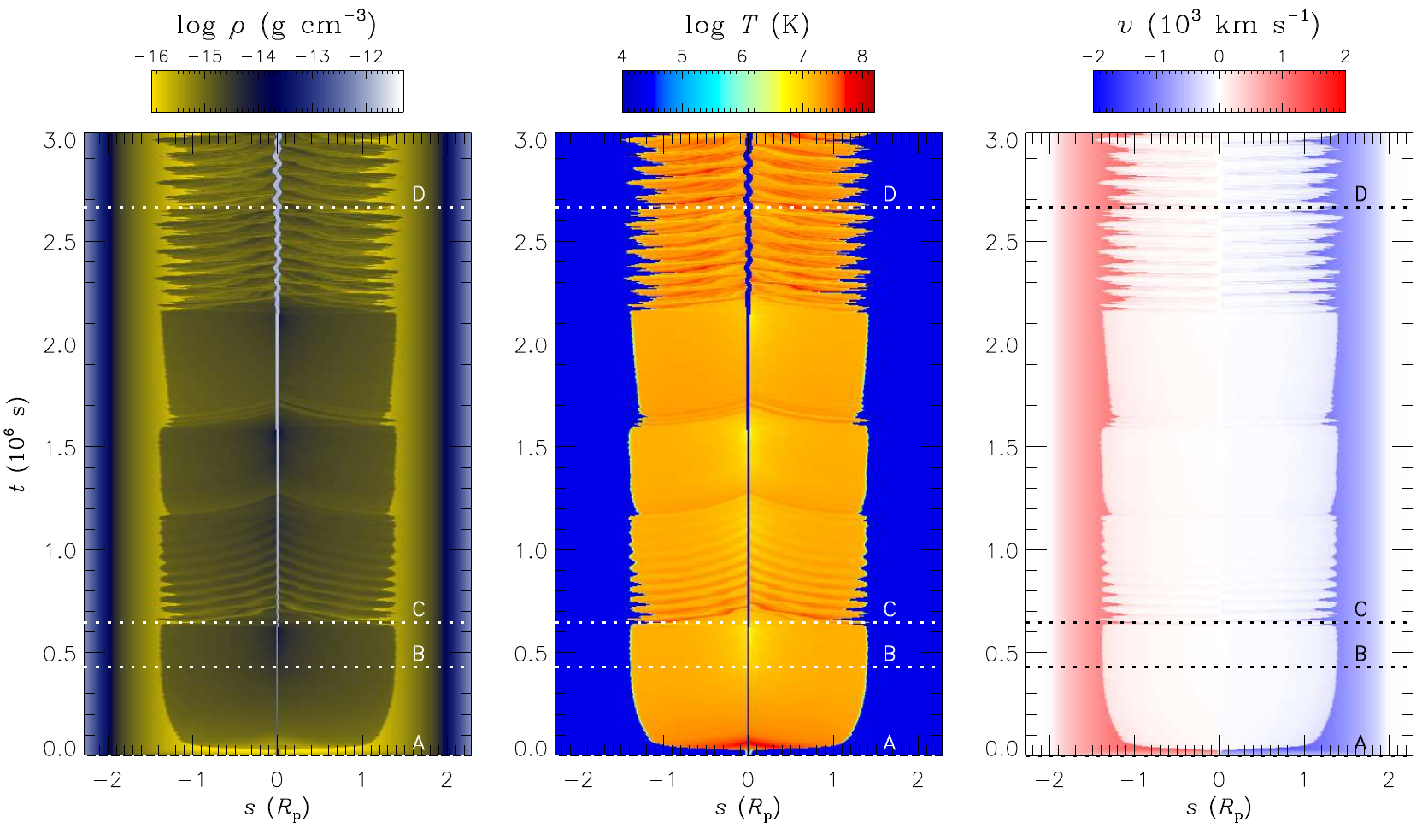}
\end{center}
\caption{The time evolution of the flow along the $(L,\pmag) =
  (2.45,90\degr)$ field line, showing the density $\rho$, temperature
  $T$, and velocity $v$ as a function of arc distance $s$ and time
  $t$. The magnetic equator is situated at $s=0$, and the two
  footpoints at $s = (\sstarN,\sstarS) = (-2.28,2.28)\,\Rpole$. The
  horizontal dotted lines, labeled alphabetically, show the locations
  of the snapshots plotted in Fig.~\ref{fig:timeslice}; the `A' line
  is situated at $t=0\,{\rm s}$.}
\label{fig:timeseries}
\end{figure*}

\begin{figure*}
\begin{center}
\includegraphics{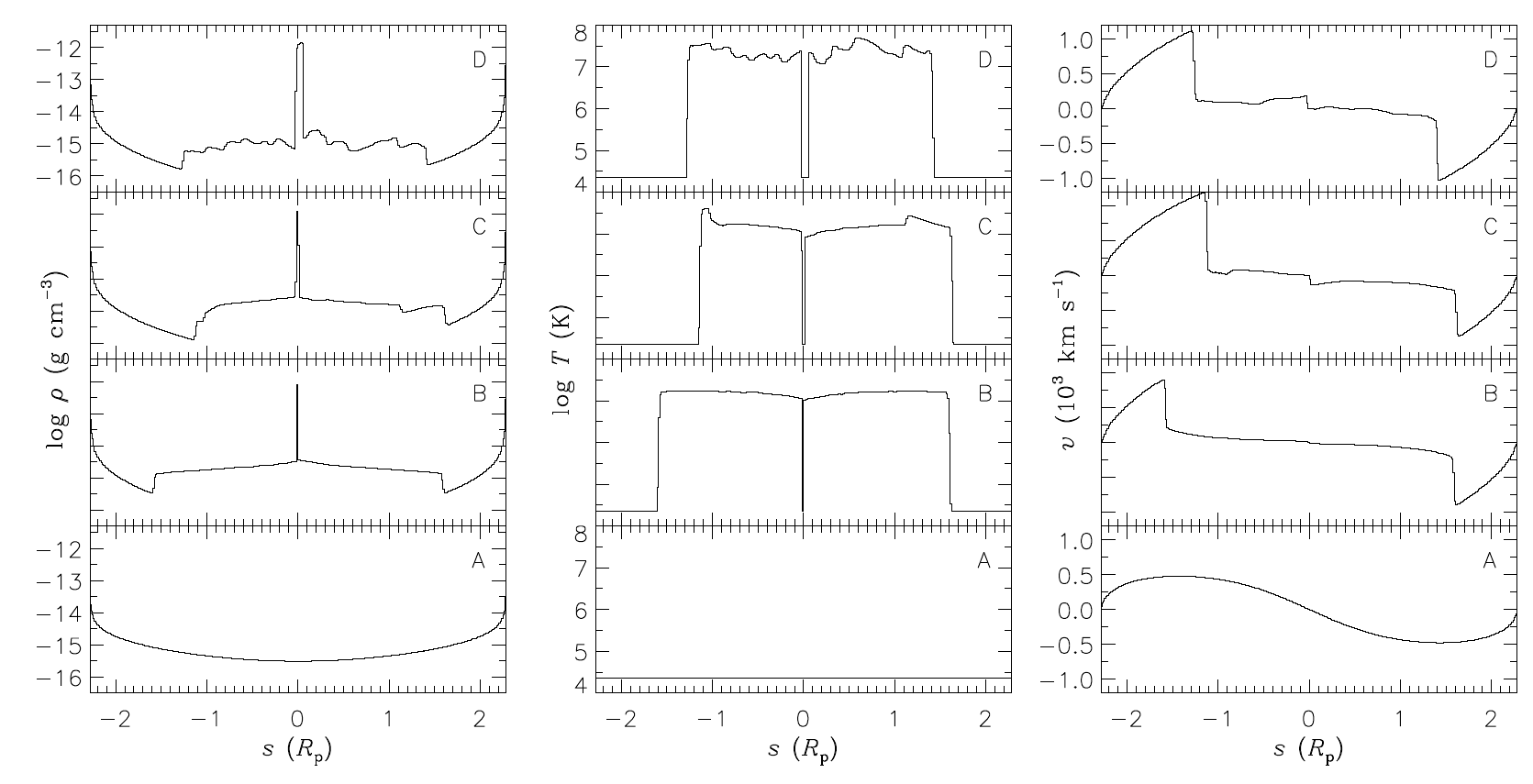}
\end{center}
\caption{Snapshots of the flow along the $(L,\pmag) =
  (2.45,90\degr)$ field line, plotting the density $\rho$,
  temperature $T$, and velocity $v$ as a function of arc distance
  $s$. The times of the snapshots are indicated in
  Fig.~\ref{fig:timeseries}.} \label{fig:timeslice}
\end{figure*}

In this section, we focus exclusively on the flow along the field line
having a magnetic shell index $\ell=20$ (cf. eqn.~\ref{eqn:req-grid}),
corresponding to $L = 2.45$, and an azimuth $\pmag = 90\degr$. With
these parameters, the field line extends out beyond the Kepler radius
$\rKep = 1.86\,\Rpole$, \emph{and} passes through the intersection
between magnetic and rotational equators. According to the RRM model
(cf. TO05), these properties should be favourable to the steady
accumulation of cooled plasma at the field-line summit. This
expectation is amply confirmed by Fig.~\ref{fig:timeseries}, which
shows the time evolution of $\rho$, $T$, and $v$ along the field line
for the first $3\,{\rm Ms}$ of the simulation. (The dynamics during
the remainder of the simulation are not significantly different from
those seen toward the end of this initial timespan). The figure should
be interpreted with the aid of Fig.~\ref{fig:timeslice}, which plots
snapshots of the flow variables at four epochs of interest.

The first, `A' snapshot shows the initial configuration described in
\S\ref{ssec:code-initial}. Because the velocity in both hemispheres is
supersonic, reverse shocks quickly form at the magnetic equator $s=0$,
and then proceed to propagate back down the field line toward each
footpoint. As wind plasma flows through these shocks it experiences an
abrupt reduction in velocity, matched by corresponding discontinuous
increases in density and temperature. The temperature jump, from $T =
\Tstar$ to $T \approx 2 \times 10^{7}\,{\rm K}$, mean that the
post-shock plasma cools initially with photon energies $k T \approx
2\,{\rm keV}$ in the X-ray range.

\begin{figure*}
\begin{center}
\includegraphics{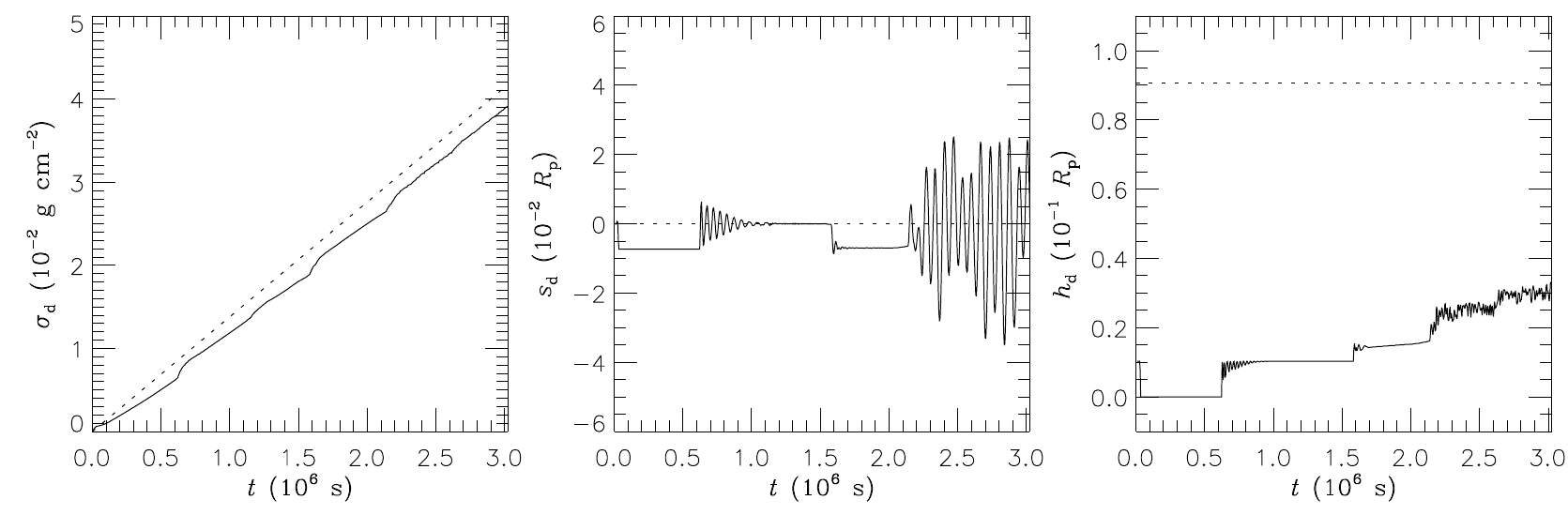}
\end{center}
\caption{The surface density \sigdisk, centroid \sdisk\ and scale
height \hdisk\ of the disk for the $(L,\pmag) = (2.45,90\degr)$ field
line, plotted as a function of time $t$. The dotted lines show the
corresponding predictions of the RRM model.}
\label{fig:moments}
\end{figure*}

The downward propagation of the shocks halts when the ram pressure of
the pre-shock plasma matches the gas pressure in the post-shock
regions. The resulting quasi-steady state, composed of standing shocks
at $s = \pm1.6\,\Rpole$ that enclose hot, post-shock cooling regions,
is shown in snapshot `B'. Situated at the centre of these regions is
the cooled plasma predicted by the RRM model. This plasma accumulates
at a local minimum of the effective potential \poteff\ (as sampled
along field lines), where it is supported in stable magnetohydrostatic
equilibrium by the centrifugal force (see TO05; see also
Appendix~\ref{app:static}). As we discuss further in
\S\ref{ssec:results-3D}, the locus formed by such potential minima
resembles an azimuthally warped disk.

To characterize the behaviour of the cool disk plasma, we briefly
digress to introduce the three moments
\begin{gather} \label{eqn:sigdisk}
\sigdisk = \frac{1}{\Aeq} \int \rho A \,\diff s\, , \\ \label{eqn:sdisk}
\sdisk = \frac{1}{\Aeq \sigdisk} \int \rho s A \,\diff s\, , \\ 
\intertext{and} \label{eqn:hdisk}
\hdisk = \left[\frac{2}{\Aeq \sigdisk} \int \rho (s - \sdisk)^{2} A
  \,\diff s\right]^{1/2}\, ,
\end{gather}
where the integrals extend over the disk regions having
$T=\Tstar$. These moments represent, respectively, the disk surface
density (projected into the magnetic equatorial plane), the disk
centroid, and the disk scale height. When applied to the 1-D
simulations, the above expressions are evaluated via finite-volume
equivalents,
\begin{gather} \label{eqn:sigdisk-finvol}
\sigdisk = \frac{1}{\Aeq} \sum \rho_{j} \,\diff V_{j}\, , \\ \label{eqn:sdisk-finvol}
\sdisk = \frac{1}{\Aeq \sigdisk} \sum \rho_{j} s_{j} \,\diff V_{j}\, , \\ \label{eqn:hdisk-finvol}
\hdisk = \left[\frac{2}{\Aeq \sigdisk} \sum \rho_{j} (s_{j} -
  \sdisk)^{2} \,\diff V_{j}\right]^{1/2}\, ,
\end{gather}
where $\diff V_{j}$ is the volume of the $j$'th zone, and $s_{j}$ the
arc distance of the zone's centre. As before, the
summations extend over disk zones having $T_{j} = \Tstar$.

Fig.~\ref{fig:moments} plots the
moments~(\ref{eqn:sigdisk-finvol}--\ref{eqn:hdisk-finvol}) for the
$(L,\pmag) = (2.45,90\degr)$ field line. The rightmost panel indicates
that during the early stages of the simulation, the disk thickness is
substantially smaller than the value $\hmin = 0.09\,\Rpole$ predicted
by the RRM model (cf. eqn.~\ref{eqn:hmin}). Indeed, up until $t =
0.6\,{\rm Ms}$ \hdisk\ is identically zero, indicating that the disk
extends over only a single zone. The reason for this confinement is
that the internal pressure of the cool disk is insufficient to support
it against the relatively high pressure of the hot plasma in the
surrounding post-shock regions; thus, the disk becomes compressed into
the smallest volume resolvable by \vhone.

This situation changes once the steady accumulation of plasma
(Fig.~\ref{fig:moments}, leftmost panel) raises the internal pressure
to a point where the disk suddenly grows to encompass a greater number
of zones. Three of these expansion events are apparent in
Fig.~\ref{fig:timeseries}, at $t = 0.6, 1.6$ and $2.1\,{\rm
Ms}$. After each event, the disk plasma undergoes oscillations back
and forth about its equilibrium position $s=0$; these are most readily
seen in the centre panel of Fig.~\ref{fig:moments} for $t > 2.1\,{\rm
Ms}$. The oscillations are accompanied by variability of the
post-shock cooling regions, involving the abrupt movement of the shock
fronts toward $s=0$, followed by a rebuilding of the cooling regions
via wind feeding. Snapshot `C' in Fig.~\ref{fig:timeslice} shows the
state of the flow immediately after an ingress of the
northern-hemisphere ($s<0$) shock; observe how this shock is
$0.45\,\Rpole$ closer to the disk than in snapshot `B'. Because the
density jump across a strong shock scales proportionally to the
pre-shock density, this movement of the shock front translates into a
reduction in the post-shock density. Thus, $\rho \approx 1.3\times
10^{-15}\,\gcmc$ just downstream of the northern shock in snapshot
`B', but $\rho \approx 0.4 \times 10^{-15}\,\gcmc$ downstream of the
shock in snapshot `C'.

The disk oscillations after the first expansion event are damped,
dying out over a timescale $\sim 0.5\,{\rm Ms}$. The damping is even
stronger for the very weak oscillations seen after the second
event. Following the third event at $t = 2.1\,{\rm Ms}$,
however, the oscillations persist at a relatively high amplitude all
the way to the end of the simulation. Snapshot `D' in
Fig.~\ref{fig:timeslice} illustrates the flow when the oscillating
disk plasma is undergoing a southward displacement ($\sdisk = 0.02\,\Rpole$).

To explore the nature of the oscillations, Appendix~\ref{app:perturb}
develops a linear analysis of the response of the disk plasma to
small-amplitude departures from stationary equilibrium. For
perturbations that conserve \sigdisk, the analysis reveals a spectrum
of normal modes having periods
\begin{equation}
P_{m} = \frac{2\pi}{\omega_{m}} = \pi \sqrt{\frac{2}{m}} \frac{\hmin}{\csoundstar} \qquad (m =
1,2,3,\ldots)\, .
\end{equation}
The oscillations seen in Figs.~\ref{fig:timeseries}--\ref{fig:moments}
are wholly consistent with the excitation of the dipole ($m=1$)
mode. In particular, the period $P_{1} = 85.5\,{\rm ks}$ predicted by
the expression above is in good agreement with the value $P =
79.2\,{\rm ks}$ measured from the strongest peak in the Fourier
transform of the \sdisk\ data for $t > 2.1\,{\rm Ms}$.

What excites the dipole oscillations? In the simulations, the
coincidence between the expansion events and the onset of oscillations
reveals that these events impart an initial `kick' to the disk
plasma. (The kick originates because the perturbation introduced by an
expansion event is usually asymmetric.)  However, some other process is
clearly responsible for maintaining and even amplifying the
oscillations, as seen after the third expansion event at $t =
2.1\,{\rm Ms}$. A likely candidate for this process is the cooling
instability discussed by \citet{CheIma1982}. These authors show that
radiative cooling of thermal plasma in the temperature range $\sim
10^{5}-10^{7}\,{\rm K}$ is linearly overstable due to the negative
temperature exponent of the cooling coefficient \coolcoeff. For the
parameters of the cooling regions of the $(L,\pmag) = (2.45,90\degr)$
field line, the periods of the fundamental and first-overtone unstable
cooling modes are $P \approx 160\,{\rm ks}$ and $P \approx 40\,{\rm
ks}$ respectively\footnote{These values are obtained from the
$\alpha=-1$ eigenfrequencies tabulated by \citet{CheIma1982}, with
$u_{\rm in} = 900\,\kms$ and $x_{s0} = -1.6\,\Rpole$.}, bracketing the
disk oscillation period $P = 79.2\,{\rm ks}$. This lends support to
the hypothesis that a coupling between the disk modes and the cooling
modes results in a global overstability, driving the disk oscillations
until non-linearity leads to saturation. We intend to investigate this
hypothesis further in a future paper.

\begin{figure}
\begin{center}
\includegraphics{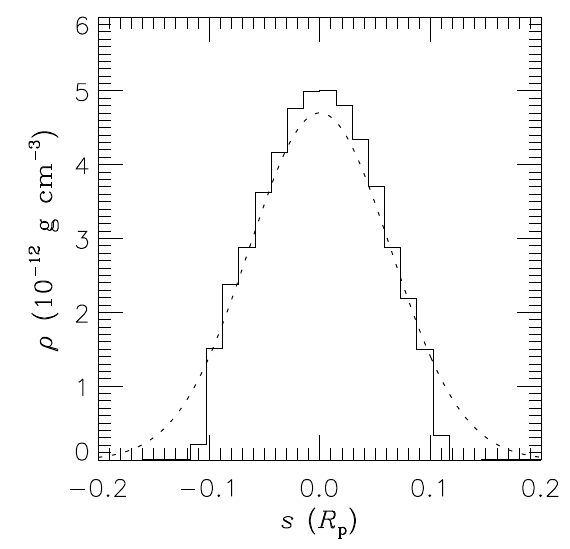}
\end{center}
\caption{The density $\rho$ in the vicinity of the cool disk, plotted
as a function of arc coordinate $s$ for the $(L,\pmag) =
(2.45,90\degr)$ field line. The solid line indicates the result from
the RFHD simulation at a time $t \approx 20.2\,{\rm Ms}$, while the dotted
curve shows the prediction of the RRM model at the same $t$.}
\label{fig:comp-1d}
\end{figure}

To bring the present section to a close, we compare the RFHD
simulation for the $(L,\pmag) = (2.45,90\degr)$ field line against the
predictions of the RRM model. For the parameters specified in
Table~\ref{tab:params}, the RRM model\footnote{See TO05, their
eqn.~(34); in evaluating the stellar-surface field tilt $\mu_{\ast}$
appearing in this equation (and in other expressions from the RRM
formalism), we take into account the stellar oblateness due to
rotation, which for simplicity was neglected in the TO05 treatment.}
predicts a rate of change $\sigdot = 1.38 \times 10^{-8}\,\gcmss$ for
the disk surface density. Given the approximations employed in
developing the model, this value is in remarkably good agreement with
the empirical value $\sigdot = 1.31 \times 10^{-8}\,\gcmss$ derived
from a linear least-squares fit to the \sigdisk\ simulation data. 

To compare the distribution of disk plasma, Fig.~\ref{fig:comp-1d}
plots the density as a function of $s$ for both model and simulation,
at a time $t \approx 20.2\,{\rm Ms}$ near the end of the simulation
chosen so that the disk displacement \sdisk\ is close to zero. Once
again, there is encouraging agreement between the RRM prediction and
the simulation result. The only significant difference is that the
wings of the Gaussian density distribution (cf.
eqn.~\ref{eqn:rrm-rho}) are truncated in the simulation, due to the
pressure of the hot plasma in the adjacent post-shock regions.

\subsection{2-D, in meridional planes} \label{ssec:results-2D}

\begin{figure*}
\begin{center}
\includegraphics{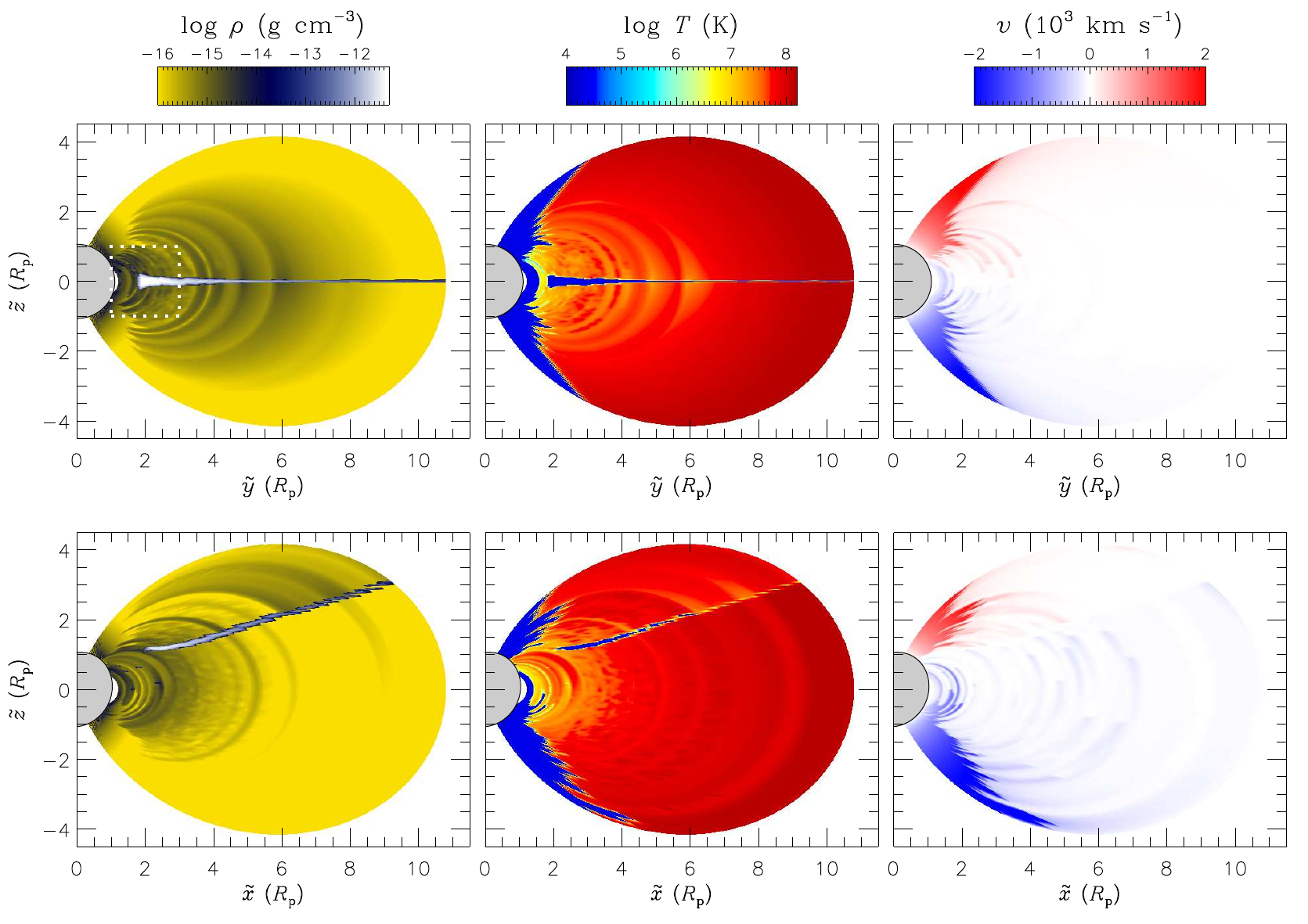}
\end{center}
\caption{The state of the flow at the end of the simulations, showing
the density $\rho$, temperature $T$, and velocity $v$ in the $\pmag =
90\degr$ (upper row) and $\pmag=0\degr$ (lower row) meridional
planes. (See Fig.~\ref{fig:geometry} for a recapitulation of the
geometry of these planes). The oblate star is indicated in grey at the
left-hand side of each image. Outside the regions threaded by
simulation field lines (i.e., for $L < 1.2$ and $L > 11.2$), the
images are left blank. The dotted rectangle in the upper left image
indicates the region shown in detail in Fig.~\ref{fig:sequence}.}
\label{fig:meridional}
\end{figure*}

We now extend the analysis to two dimensions, by considering 1-D RFHD
simulations all lying in the same meridional plane.
Fig.~\ref{fig:meridional} shows 2-D images of the state of the flow at
the end of the simulations, for the $\pmag=90\degr$ meridional
plane. (The figure also shows images for the $\pmag=0\degr$ plane, but
we defer discussion of these data until later).  The cool disk of
accumulated plasma is clearly seen along the \ymag\ axis in the
$\pmag=90\degr$ plane, surrounded by hot ($T \sim 10^{7}-10^{8}\,{\rm
K}$) post-shock cooling regions. The disk does not extend all the way
to the star, but is instead truncated at a radius $r =
1.8\,\Rpole$. Inside this radius, the centrifugal force is not strong
enough to support plasma against the inward pull of gravity, and no
accumulation occurs. Although dense knots of plasma are formed close
to the star by compression between opposing wind streams, these knots
quickly slide down the magnetic field toward one or the other of the
footpoints. 

\begin{figure*}
\begin{center}
\includegraphics{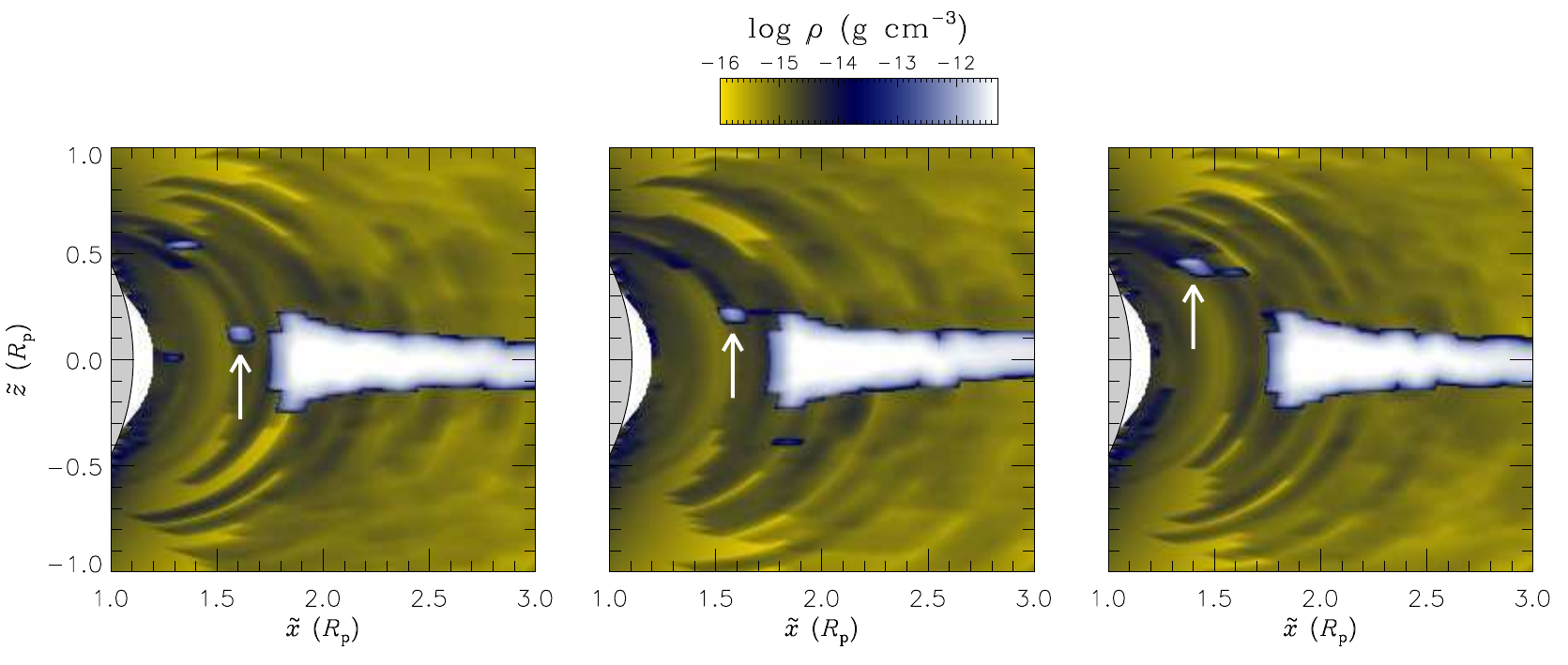}
\end{center}
\caption{Snapshots of the density $\rho$ in the inner parts of the
$\pmag = 90\degr$ meridional plane (see the dotted rectangle in
Fig.~\ref{fig:meridional}), at a time $t \approx 20.4\,{\rm Ms}$ near
the end of the simulations (left), and then at increments of one
quarter (centre) and one half (right) of a rotation cycle later. The
white arrows indicate the location of the dense knot discussed in the
text.  As in Fig.~\ref{fig:meridional}, the star is indicated in grey
at the left-hand side of each image.}
\label{fig:sequence}
\end{figure*}

Figure~\ref{fig:sequence} illustrates this fallback process in a
sequence of density snapshots, showing the evolution of the inner
parts of the magnetosphere shortly before the end of the
simulations. The white arrow in the leftmost panel indicates a knot
that has formed at $(\ymag,\zmag) = (1.6,0.1) \,\Rpole$. In the centre
panel, the knot has migrated further into the northern magnetic
hemisphere, and in the rightmost panel begun its descent to the
stellar surface. This figure is reminiscent of the infalling plasma
seen in the MHD simulations by \citet[][their fig.~4]{udDOwo2002}; in
fact, the only phenomenological difference lies in the scale of the
knots. Cross-field coupling in the MHD simulations allows coherence
between the flow on adjacent field lines. However, this coupling is
absent in the RFHD simulations (due to the neglect of the polar
velocity derivative when calculating the radiative acceleration; see
\S\ref{ssec:rfhd-rad}), and the scale of the knots is set therefore by
the $L$ grid spacing (cf. eqn.~\ref{eqn:req-grid}).

The absence of cross-field coupling is also ultimately responsible for
the significant amount of structure seen in the post-shock cooling
regions. If disk oscillations are excited on one particular field
line, but are absent from an adjacent field line, then discontinuities
in the cross-field direction appear in the flow properties, resembling
a reversed letter `C'. Thus, for instance, the $(L,\pmag) =
(3.88,90\degr)$ field line in Fig.~\ref{fig:meridional} has a
significantly lower density than its neighbours because disk
oscillations are excited on this field line, but not on the adjacent
ones. As with the fallback, we expect in reality that cross-field
coupling will tend to smear out such sharp discontinuities. One
possible approach to including this coupling in RFHD simulations is
discussed in \S\ref{sssec:discuss-limit-cross}.

Turning now to the temperature data in Fig.~\ref{fig:meridional}, a
gradient can be seen across the post-shock regions, with the outer
parts being up to an order of magnitude hotter ($T\sim 10^{8}\,{\rm
K}$) than those near the star ($T\sim 10^{7}\,{\rm K}$). This gradient
arises from two distinct effects.  First, field lines having larger
magnetic shell parameter $L$ undergo a greater area divergence, and
are therefore characterized by lower plasma densities and faster flow
velocities \citep[see][for a more in-depth discussion of this
effect]{OwoudD2004}. This naturally leads to a tendency for larger-$L$
field lines to experience hotter post-shock temperatures.

Second, in a process first conjectured by \citet{BabMon1997a}, the
post-shock plasma can be pushed to even higher temperatures by the
action of the centrifugal force. To accumulate onto the cool disk, the
plasma must first descend to the bottom of the effective potential
well (\S\ref{ssec:results-1D}). The consequent release of centrifugal
potential energy heats the plasma, by an amount that ultimately
depends on the distance from the rotation axis. In the $\pmag=90\degr$
data shown in Fig.~\ref{fig:meridional} this effect raises the plasma
temperature on the outer, $L = 11.2$ field line from $T = 1.1 \times
10^{8}\,{\rm K}$ at the shocks, to $T = 1.8 \times 10^{8}\,{\rm
K}$ adjacent to the disk.

The centrifugal heating in the post-shock regions competes against
cooling by atomic and inverse Compton processes (cf.
\S\ref{ssec:code-cool}). Initially, the centrifugal effect dominates,
because plasma densities are so low that atomic cooling (a $\rho^{2}$
process; see eqn.~\ref{eqn:coolat}) is exceedingly
inefficient. However, this situation is subsequently reversed; as
plasma moves downstream of the shocks, the density eventually becomes
high enough for atomic processes to cool it rapidly, in thin layers on
either side of the disk (see \S\ref{ssec:results-3D}). During this
process, inverse Compton cooling is relatively unimportant; at low
densities it is more efficient than atomic cooling, but it never
becomes the dominant term on the right-hand side of the energy
conservation equation~(\ref{eqn:hydro-energy}).

Although the foregoing discussion focuses on the flow in the
$\pmag=90\degr$ meridional plane, it generally applies to other
meridional planes. However, one significant exception concerns the
location of the cool disk. The $\pmag=0\degr$ images in
Fig.~\ref{fig:meridional} reveal that the disk is not symmetric about
the magnetic axis, as one might presume from considering the
$\pmag=90\degr$ images on their own. Rather, the disk is warped in the
\emph{azimuthal} direction. To explain this notion, let \tmagdisk\
denote the centroid magnetic colatitude of the disk, obtained by
setting $s=\sdisk$ in eqn.~(\ref{eqn:s}) and solving for \tmag. Then,
in a given meridional plane the disk lies approximately along a
straight ray emanating from the origin, with \tmagdisk\ remaining
constant as $L$ is varied. However, \tmagdisk\ changes from plane to
plane, resulting in azimuthal warping; for instance, $\tmagdisk
\approx 70\degr$ in the $\pmag=0\degr$ meridional plane, whereas
$\tmagdisk \approx 90\degr$ for the $\pmag=90\degr$ case. This warping
is one of the key predictions of the RRM model, and its origin is
discussed in greater detail by TO05.

In addition to revealing the disk warping, the $\pmag=0\degr$ images
in Fig.~\ref{fig:meridional} also illustrate a novel flow
phenomenon. On a number of the innermost ($L \lesssim 1.8$) field
lines, the northern reverse shock propagates back down to the stellar
surface. This leads to a siphon configuration, in which plasma flows
unidirectionally from the southern magnetic footpoint to the
corresponding northern footpoint. (The direction of the flow is set by
the relative location of the rotation axis; in the $\pmag=180\degr$
meridional plane, north-to-south siphon flows occur). Because the
southern reverse shock remains intact, part of the siphon flow is
composed of shock-heated plasma at $T \approx 5\times10^{6}\,{\rm K}$;
this plasma is relatively dense due to its proximity to the star, and
therefore makes a significant contribution to the soft X-ray emission
from the magnetosphere (see \S\ref{ssec:results-3D}).

\begin{figure}
\begin{center}
\includegraphics{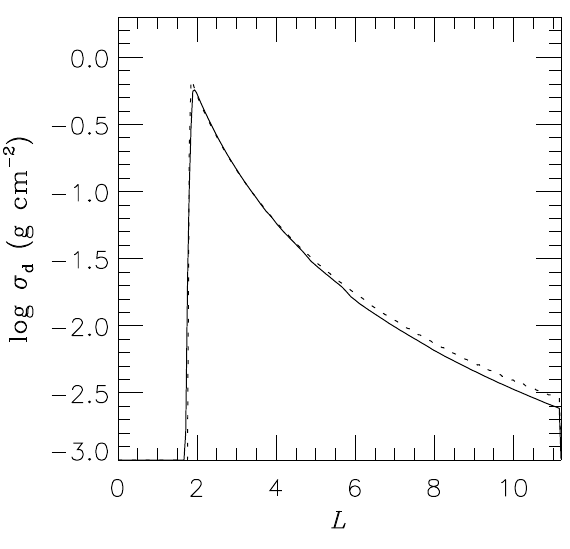}
\end{center}
\caption{The disk surface density \sigdisk\ in the $\pmag = 90\degr$
meridional plane, plotted as a function of magnetic shell parameter
$L$. The solid line shows the final state at the end of the RFHD
simulations, while the dotted line indicates the corresponding
prediction of the RRM model.} \label{fig:comp-2d}
\end{figure}

As with the 1-D analysis in \S\ref{ssec:results-1D}, we bring the
present section to a close by comparing the RFHD simulation results
against the predictions of the corresponding RRM
model. Fig.~\ref{fig:comp-2d} plots the disk surface density \sigdisk\
in the $\pmag=90\degr$ meridional plane, at the end of the simulations
and for the model. The agreement between the two is once again very
encouraging, especially in the innermost regions of the disk. In the
outer regions, the simulations predict a rather smaller surface
density ($\sim 10$--25\%) than the RRM model. This modest discrepancy
appears correlated with the observation that, for field lines having
$L \gtrsim 6$, the disk remains confined to a single zone -- no
expansion events occur on these field lines over the duration of the simulations.

\subsection{Full 3-D model} \label{ssec:results-3D}

\begin{figure*}
\begin{center}
\includegraphics{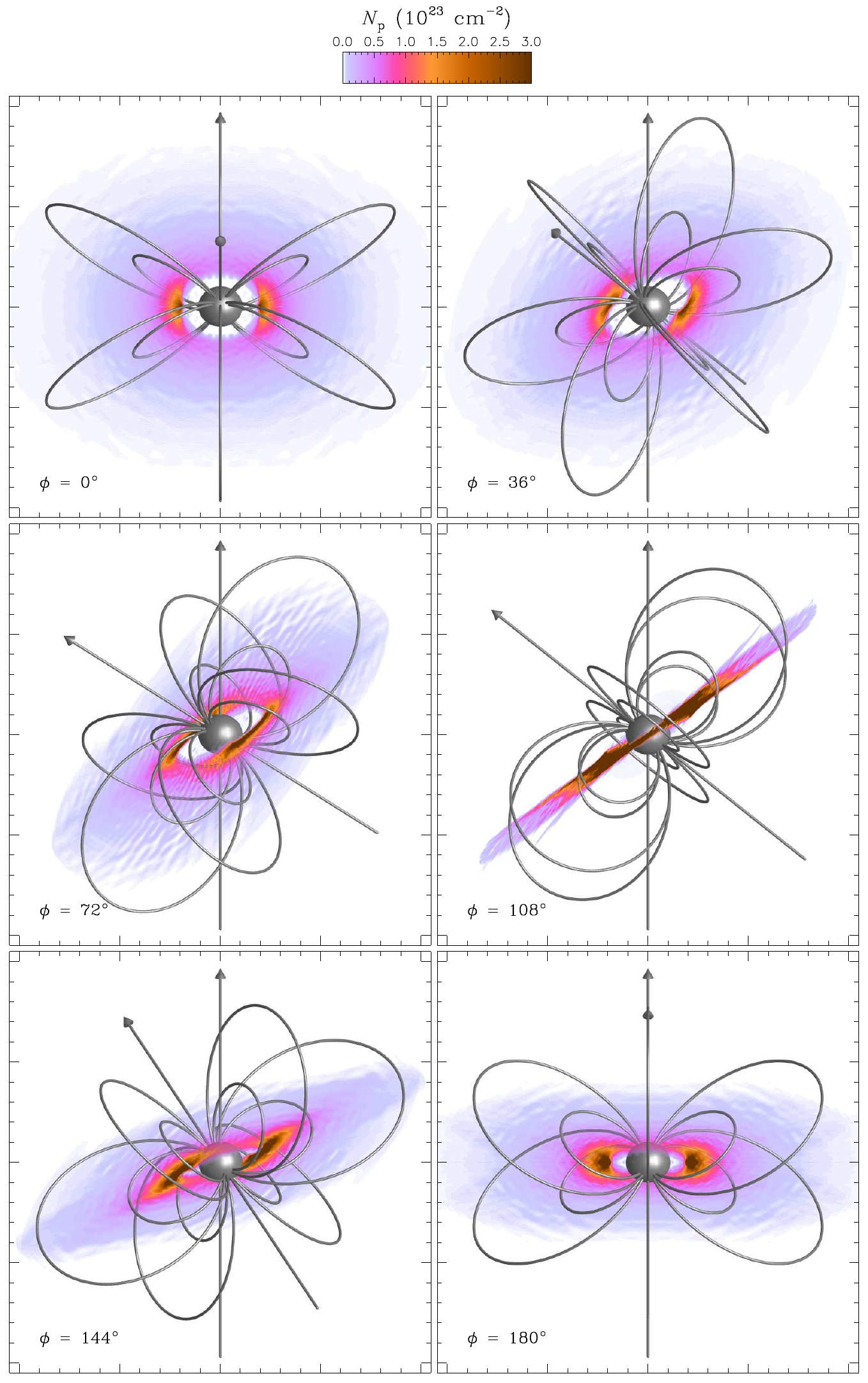}
\end{center}
\caption{The proton column density \Np\ at the end of the simulations,
viewed from six different rotational azimuths \prot. The oblate star
is shown in grey at the centre of each panel; arrows indicate the
magnetic and rotation axes (the latter being the vertical, fixed one),
and the curved arcs show field lines having magnetic shell parameters
$L=5,10$ and magnetic azimuths
$\pmag=0\degr,60\degr,120\degr,\ldots,300\degr$.}
\label{fig:column}
\end{figure*}

Having examined the results of the RFHD simulations for an individual
field line, and for field lines lying in meridional planes, we now
turn to the complete, three-dimensional picture. Fig.~\ref{fig:column}
shows images of the proton column density \Np\ at the end of the
simulations, viewed from six equally spaced rotational azimuths spanning
the range $0\degr \leq \prot \leq 180\degr$ (the remaining interval
$180\degr \leq \prot \leq 360\degr$ is mirror-symmetric through the
vertical axis). Following \citet{Tow2005}, a viewing inclination
$i=75\degr$ with respect to the rotation axis is adopted. The column
density is calculated from ray integrals of the form
\begin{equation}
\Np = \int \np \,\diff\zimg\, ,
\end{equation}
where \zimg\ is the perpendicular distance to the image plane, and the
proton number density \np\ was defined in eqn.~(\ref{eqn:np}). The
mass density $\rho$ appearing in this latter equation is calculated
using trilinear interpolation from the $(s,L,\pmag)$ field-line
coordinate system onto the $(\ximg,\yimg,\zimg)$ Cartesian image
coordinate system. Along rays that intersect the star, the
integral is truncated at the stellar surface.

The figure clearly illustrates the three-dimensional structure of the
dense, co-rotating disk discussed in previous sections. (The plasma in
the wind and post-shock regions is effectively invisible, since it is
orders-of-magnitude less dense than that comprising the disk.) This
disk possesses three important characteristics predicted by the RRM
model. First, it has an average inclination that lies somewhere
between the magnetic and rotational equatorial planes. Second, it
exhibits a hole at its centre. Third, the surface density across the
disk is non-uniform, with the plasma concentrated into two elongated
`clouds' situated at the intersection between magnetic and rotational
equators. These clouds are best seen in the $\prot = 180\degr$ panel
of the figure. In the case of \sorie, the existence of such clouds has
been inferred from \Halpha\ measurements
\citep[e.g.,][]{GroHun1982,Bol1987}, and \citet{Tow2005} have
demonstrated how the clouds are simultaneously responsible for the
distinctive spectroscopic and photometric variability exhibited by the
star.

\begin{figure*}
\begin{center}
\includegraphics{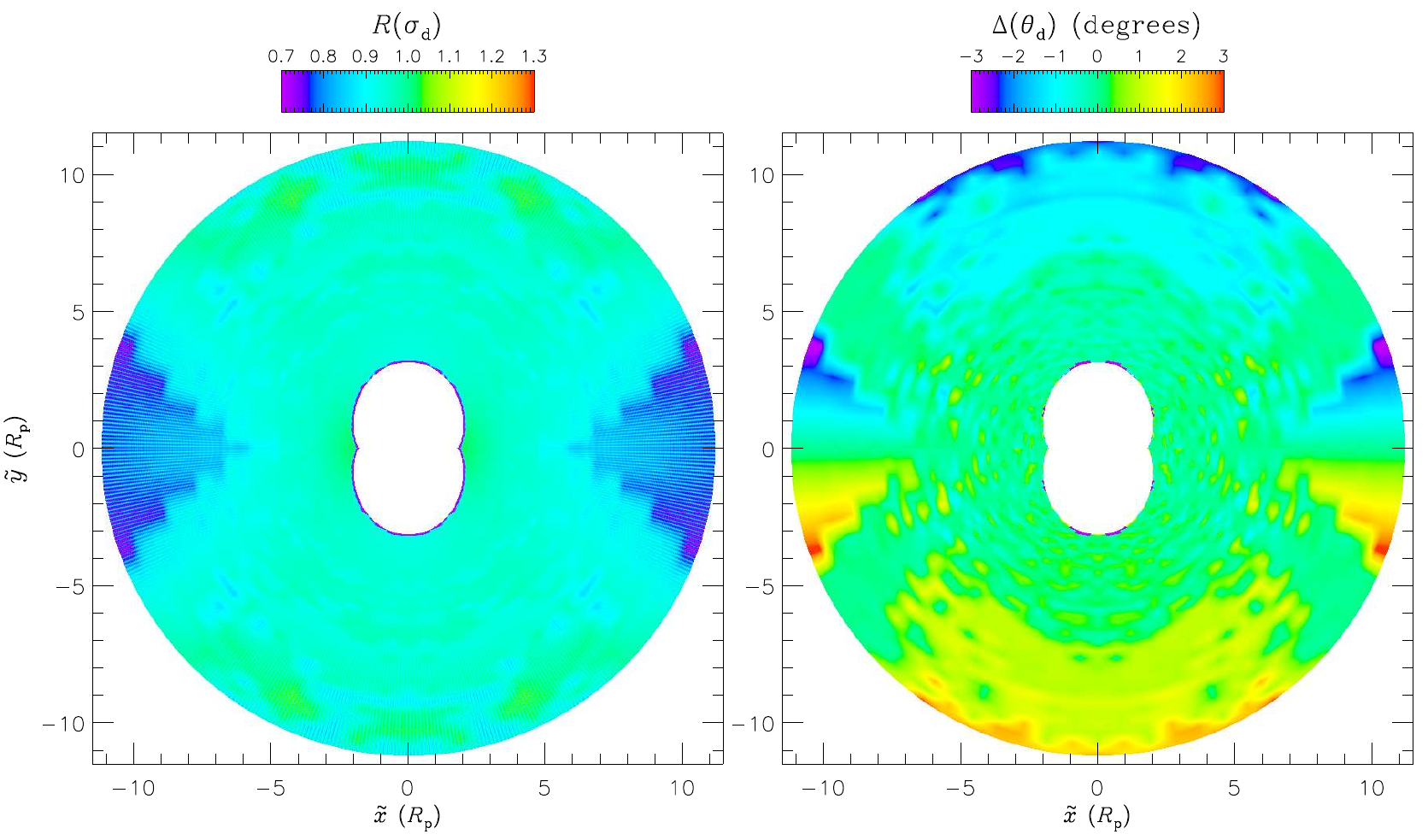}
\end{center}
\caption{The surface density ratio (left) and centroid colatitude
difference (right) between the RFHD simulations and the RRM model,
across the magnetic equatorial (\xmag--\ymag) plane.} \label{fig:comp-3d}
\end{figure*}

To explore further the degree of correspondence with the RRM model,
Fig.~\ref{fig:comp-3d} plots the quantities
\begin{equation}
\Rsigma = \frac{\sigdiskrfhd}{\sigdiskrrm}
\end{equation}
and
\begin{equation}
\Dtmag = \tmagdiskrfhd - \tmagdiskrrm
\end{equation}
in the $\xmag-\ymag$ magnetic equatorial plane, where the subscripts $_{\rm
RFHD}$ and $_{\rm RRM}$ denote values obtained from the RFHD
simulations and the RRM model, respectively. These quantities
represent the ratio of disk surface densities \sigdisk, as defined by
eqn.~(\ref{eqn:sigdisk}), and the difference in the disk centroid
colatitudes \tmagdisk\ introduced in \S\ref{ssec:results-2D}.

The figure reveals that once again the agreement between simulations
and model is good. The maximum differences in surface density are
below the 25\% level, and -- as already discussed -- are strongly
correlated with those disk regions that have not undergone any
expansion events. The differences in centroid colatitude angle do not
rise above a few degrees; they tend to be most positive around
$\pmag=0\degr$, and most negative around $\pmag=180\degr$, so the RFHD
disk is slightly closer to the magnetic equatorial plane than the RRM
disk. Based on these findings, we can conclude that the disk plasma
distribution predicted by the RRM model furnishes a close
approximation to the distribution determined via the physically
more-sophisticated (yet computationally more expensive) RFHD approach.

\begin{figure*}
\begin{center}
\includegraphics{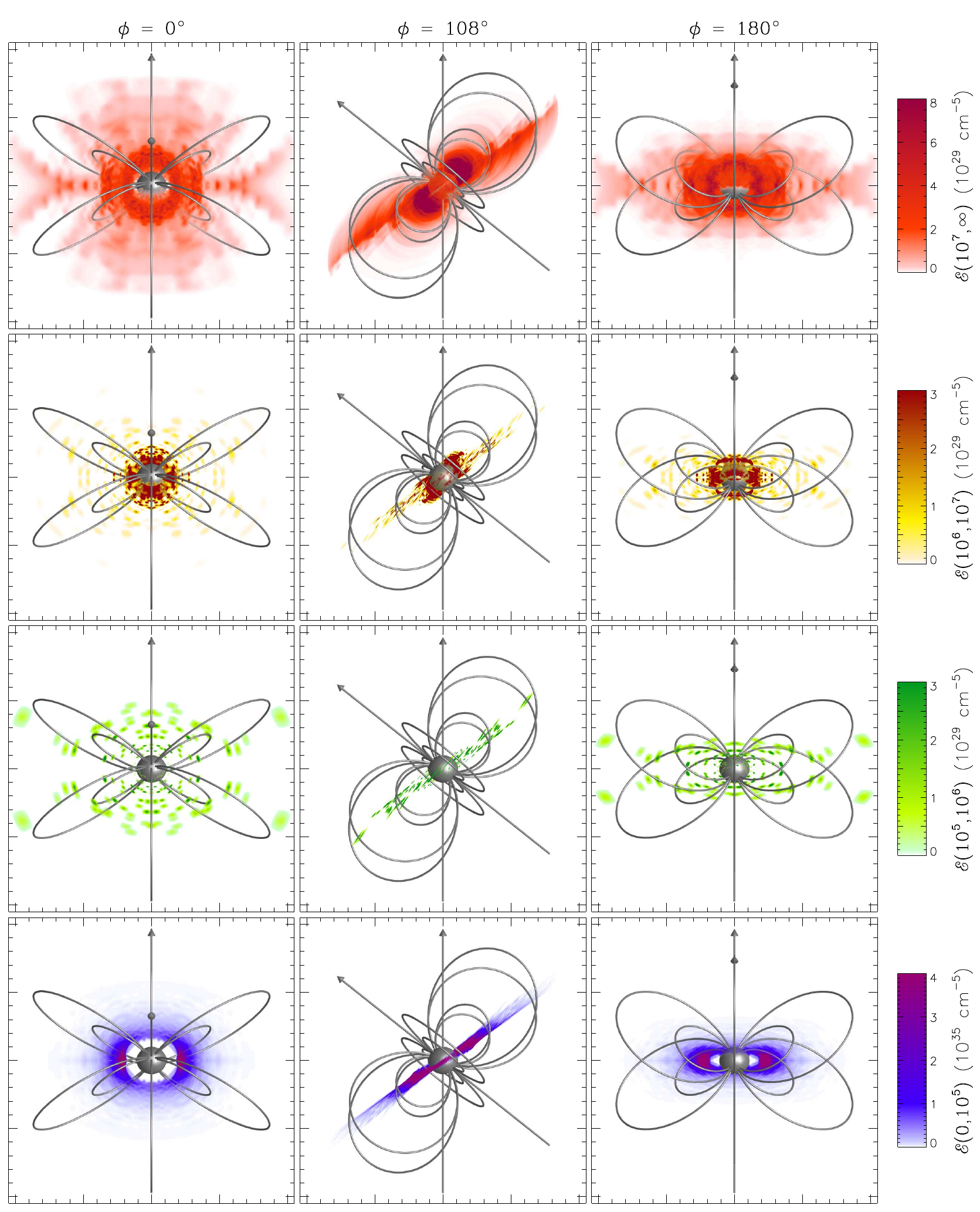}
\end{center}
\caption{The emission measure density $\emd(\Tlo,\Thi)$, viewed from
three different rotational azimuths \prot, and for four temperature
ranges $(\Tlo,\Thi)$ corresponding (top-to-bottom) to hard X-ray, soft
X-ray, EUV, and optical/UV emission. The star, axes, and field lines
are the same as in Fig.~\ref{fig:column}.} \label{fig:emission}
\end{figure*}

Of course, the significant caveat here is that the RRM model is limited
to consideration only of the cool plasma in the disk. Whereas, the RFHD
simulations also encompass the hot post-shock plasma responsible, for
instance, for magnetospheric X-ray emission. To illustrate the spatial
and thermal distribution of this plasma, we introduce a
two-temperature emission measure density (EMD),
\begin{equation}
\emd(\Tlo,\Thi) = \int_{\Tlo}^{\Thi} \left[ \int \nel \np \delta(T -
  T') \,\diff\zimg \right] \,\diff T'\, ,
\end{equation}
where $\delta()$ is the Dirac delta function, and \nel\ is the
electron number density (cf. eqn.~\ref{eqn:nel}). The EMD
characterizes the radiative recombination emission by plasma in the
temperature range $\Tlo < T < \Thi$. Fig.~\ref{fig:emission} shows
images of \emd\ at the end of the simulations, for the `face on'
($\prot=0\degr,180\degr$) and `edge-on' ($\prot=108\degr$) viewing
aspects, and for temperature ranges that correspond to thermal
emission at optical/UV ($T < 10^{5}\,{\rm K}$), extreme UV
($10^{5}\,{\rm K} < T < 10^{6}\,{\rm K}$), soft X-ray ($10^{6}\,{\rm
K} < T < 10^{7}\,{\rm K}$) and hard X-ray ($T > 10^{7}\,{\rm K}$)
energies.

The optical/UV images confirm the concentration of \Halpha-emitting
plasma into two clouds. However, at higher temperatures a different
distribution emerges. In the EUV range, the plasma is situated
primarily in thin cooling layers on either side of the disk. Because
the cooling from $10^{6}\,{\rm K}$ to $10^{5}\,{\rm K}$ is so
efficient (on account of the large number of metal lines available),
these layers are under-resolved in the simulations, and appear
fragmented into many small islands of emission. Similarly fragmented
cooling layers are seen in the outer parts of the soft X-ray EMD images,
but the preponderance of the emission in this $10^{6}\,{\rm K} < T <
10^{7}\,{\rm K}$ range comes from a toroidal belt surrounding the
star. This belt is primarily composed of the hot, dense plasma
associated with the siphon flows discussed in \S\ref{ssec:results-2D}
(and see also Fig.~\ref{fig:meridional}).

The soft X-ray belt is enclosed by a torus of hard X-ray emission,
seen in the $T > 10^{7}\,{\rm K}$ images to extend out to about
4\,\Rpole. Beyond this radius, an asymmetry develops on either side of
the disk, with those field lines that pass near the rotational poles
showing much weaker emission than those that avoid the poles. This
asymmetry is most evident in the $\prot=108\degr$ (edge-on) panel;
observe how the emission is strongest in the northern magnetic
hemisphere on the left-hand side of the image, but vice-versa on the
right-hand side. The origin of the asymmetry can be seen in the
$\pmag=0\degr$ images of Fig.~\ref{fig:meridional}; note the
significant difference in the density of the post-shock regions on
either side of the disk, which translates directly into the \emd\
asymmetry. The density difference itself arises as a result of the
positioning of the shock fronts; the disk in the $\pmag=0\degr$
meridional plane is situated in the northern magnetic hemisphere, and
the shock fronts in this hemisphere are closer to the stellar surface,
implying a higher density in the post-shock region.

\begin{figure}
\begin{center}
\includegraphics{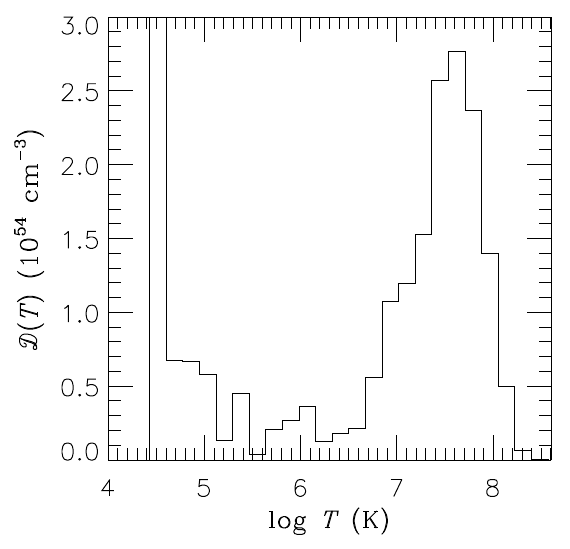}
\end{center}
\caption{The differential emission measure distribution \dem, plotted
as a function of temperature $T$ for the $\prot=0\degr$ case shown in
Fig.~\ref{fig:emission}. The ordinate scale is chosen to emphasize the
emission at EUV and X-ray energies; the truncated peak at low
temperatures extends up to $\approx 2 \times 10^{60}\,{\rm cm^{-3}}$.}
\label{fig:dem}
\end{figure}

To complete our discussion of the magnetospheric emission, we consider
the differential emission measure (DEM) distribution
\begin{equation}
\dem(T) = \frac{\diff}{\diff \ln T} \left[ \int\int \emd(0,T) \,\diff
  \ximg \,\diff\yimg \right]\, .
\end{equation}
Fig.~\ref{fig:dem} plots \dem\ as a function of temperature,
for the $\prot=0\degr$ case shown in Fig.~\ref{fig:emission}.  The
figure reveals an essentially bimodal DEM distribution, with a narrow
peak at $T = 22,500\,{\rm K} = \Tstar$ corresponding to the cool disk
plasma, and a broad peak centred at $T \approx 4 \times 10^{7}\,{\rm
K}$ associated with the extended regions of hard X-ray emission seen
in the topmost images of Fig.~\ref{fig:emission}. These
regions are not subject to any appreciable occultation by the star,
and as a consequence there is almost no change to the X-ray peak seen
in Fig.~\ref{fig:dem} as the rotational azimuth \prot\ is varied.


\section{Discussion} \label{sec:discuss}

The analysis in preceding sections is largely directed toward
comparing the results from the RFHD simulations against the
corresponding predictions of the RRM formalism. This comparison
confirms that the two complementary frameworks for modeling
massive-star magnetospheres are in good agreement. However, the RFHD
approach has a far broader scope than the RRM model, and we have only
begun to explore its potential applications. Our preliminary
investigation of a \sorie-like star has already revealed a variety of
novel phenomena, such as disk oscillations and centrifugal heating.
In \S\ref{ssec:discuss-obs} we discuss the prospects for the next
logical step of comparing RFHD simulations against observations of
magnetic stars. In \S\ref{ssec:discuss-limit}, we review the
limitations of the RFHD approach, and suggest how these might be
overcome in future studies. First, however, we examine the
relationship between RFHD and previous work.

\subsection{Relation to previous work} \label{ssec:discuss-previous}

The RFHD approach presented in this paper has allowed (to our
knowledge) the first-ever time-dependent, three-dimensional simulation
of a massive-star magnetosphere.  In developing the new approach, we
have drawn extensively on previous investigations of stellar
magnetospheres. From the perspective of disk accumulation, these
include the RRM model (TO05), and the earlier rigid-field models
advanced by \citet{MicStu1974} and \citet{Nak1985}. From the
perspective of wind dynamics, the key narrative has been the MCWS
paradigm of \citet{BabMon1997a}.

Indeed, the modeling strategy employed by \citet{BabMon1997a} directly
foreshadows the RFHD approach, and a comparison between the two is
appropriate. These authors also adopt a rigid-field ansatz, and
likewise cast the flow in terms of independent 1-D hydrodynamical
problems that include the effects of rotation and radiative
cooling. However, they restrict their analysis to the 2-D axisymmetric
case of an aligned dipole ($\beta=0\degr$), and instead of
constructing global solutions they treat the wind, post-shock and disk
regions separately, combining them in a \emph{post hoc} step based on
matching conditions. Moreover, rather than using a hydrodynamical code
they treat the flow in the wind and post-shock regions by stationary
integration of the time-independent momentum equation. This rules out
any possibility of simulating dynamical phenomena such as disk
oscillations or the fallback of material close to the star.

For these reasons, we regard the RFHD approach as a significant
advance beyond the MCWS model that augured it. What of other
treatments?  As discussed in the introduction, MHD simulations
\citep[e.g.,][]{udDOwo2002} tend to be impractical when field lines
become almost rigid. Therefore, the direct overlap between MHD and
RFHD is naturally limited -- although this may change if the RFHD
approach can successfully be extended to open field topologies, as is
discussed below in \S\ref{sssec:discuss-limit-open}. Even for stars
accessible to MHD simulation, the RFHD approach will retain some
advantage due to its relatively low computational costs.

\subsection{Comparison with observations} \label{ssec:discuss-obs}

In presenting the results from the 3-D magnetosphere model
(\S\ref{sec:results}), we avoid any specific comparison with
observations. This reflects the focus of the paper towards introducing
the RFHD approach and exploring the physical processes at work in
massive-star magnetospheres. Here, we outline those specific areas
where we believe a confrontation between models and observations will
prove most fruitful. However, apart from brief remarks on the
correspondence of our results to recent and historical observations,
we defer the detailed quantitative modeling to subsequent papers.

\citet{Tow2005} have already found that the RRM model furnishes a good
agreement to the \Halpha\ and photometric variability of the
\sorie. Since the RFHD simulations predict a disk plasma distribution
similar to the RRM model (cf.~\S\ref{ssec:results-3D}), the
improvements brought by the simulations -- at optical wavelengths --
are likely to be incremental. Nevertheless, the radiative transfer
treatment adopted by \citet{Tow2005} is at the simplest possible
level, and lacks a realistic treatment of atomic physics. Thus, there
is ample room for progress in this area.

The scope for progress is greater at other wavelengths, however. In
their \chandra\ survey of M17, \citet{Bro2007} uncovered a population
of B0-B3 stars characterized by hard (up to $\sim 5\,{\rm keV}$) X-ray
emission. The DEM distributions presented in \S\ref{ssec:results-3D}
are generally consistent with this level of hardness, suggesting a
magnetic wind-shock origin for the B stars' X-rays. To test this
hypothesis at a qualitative level, the RFHD simulations can be
combined with standard emission codes such as \apec\ \citep{Smi2001},
to synthesize spectra for direct comparison against the observations.

Such modeling should also help to understand the X-ray emission of
known-magnetic Bp stars. Some of these are notable on account of the
absence of any significant X-ray detections \citep[see,
e.g.][]{Dra1994,CzeSch2007}.  In the case of \sorie\ X-rays \emph{are}
seen, but they show an ambiguous character: a two-temperature fit to
\chandra\ measurements gives $kT \approx 0.7,2.2-3.8\,{\rm keV}$
\citep{Ski2004}, whereas a corresponding fit to \xmm\ data finds $kT
\approx 0.3,1.1\,{\rm keV}$ \citep*{San2004}. The RFHD simulations
based loosely on the parameters of \sorie\ predict a non-varying and
rather harder spectrum, with a DEM distribution peaking around $kT
\approx 3.5\,{\rm keV}$ (cf. Fig.~\ref{fig:dem}). This
harder-than-observed spectrum could be a consequence of our neglect of
thermal conduction (cf. \S\ref{ssec:rfhd-cool}).  Alternatively, it
could be that the mass-loss rate adopted in the simulations is too
high. (A larger \Mdot\ leads to a higher-density wind, which in turn
pushes the shocks further from the star where the flow velocity is
higher.) In this respect, we note that there is mounting evidence from
a number of independent diagnostics that mass-loss rates of OB stars
have historically been overestimated \citep[see, e.g.][and references
therein]{SmiOwo2006}.

At radio wavelengths, the emission properties of magnetic Bp stars are
rather more consistent than for X-rays. However, the physical
mechanisms responsible remain somewhat controversial
\citep[e.g.,][]{Dra1998}. In a recent paper, \citet{Tri2004} advance a
3-D model incorporating gyrosynchrotron emission from
mildly-relativistic electrons accelerated in an equatorial current
sheet. This current sheet lies outside the inner, rigid-field regions
of the magnetosphere, and cannot be modeled directly using the RFHD
approach (although see \S\ref{sssec:discuss-limit-open}). However,
\citet{Tri2004} suggest that the thermal plasma trapped in the
post-shock cooling regions of the inner magnetosphere plays an
important role in modulating (via free-free absorption) the radio
emission. The RFHD approach will prove useful in testing this idea.

At UV wavelengths, there is good potential for headway to be made. A
number of magnetic massive stars were observed extensively by \iue\
\citep[see][and references therein]{SmiGro2001}, leading to empirical
models for equatorially focused wind streams
\citep[e.g.,][]{ShoBro1990} that prefigured the MCWS model. In the
case of \sorie\ and similar He-strong stars, \citet{SmiGro2001} find
that the strengths of UV lines are consistent with absorbing column
densities of $10^{22}-10^{23}\,\cms$. Allowing for the fact that these
authors assumed a covering factor of 100\%\ (whereas magnetospheric
disks are in fact closer to 10\%), such values are consistent with the
$\Np \approx 3 \times 10^{23}\,\cms$ column densities found in the
RFHD simulations. The task now is to investigate whether the
simulations can reproduce the detailed absorption profiles measured by
\iue.

On a more speculative note, the RFHD approach may be able to shed some
light on particle acceleration processes around massive
stars. \citet{Bel1978a,Bel1978b} and \citet{BlaOst1978} independently
suggested that Fermi acceleration of particles in astrophysical shocks
could explain the cosmic-ray energy spectrum up to the `knee' at $\sim
10^{6}-10^{7}\,{\rm GeV}$. Typically, supernova remnants are
considered the most likely locations for this to occur
\citep{Sta2004}. However, it seems possible that the circumstellar
environments of strong-field massive stars could also be sites for
particle acceleration. \citet{BabMon1997a} suggest that the highly
sub-Alfv\'{e}nic nature of the magnetically channeled wind shocks (due
to the near-rigid field) means that the \emph{second-order} Fermi
process should be effective. They claim that relativistic electrons
produced in this manner can explain the radio emission from magnetic
massive stars, without the need for the current sheets invoked by
\citet{Tri2004}.

We conjecture that the same mechanism operating on protons and/or ions
(which are far less sensitive to inverse Compton losses than
electrons) might ultimately lead to the production of energetic
$\gamma$ rays. With recent advent of high-sensitivity ground-based
Cherenkov telescopes such as \hess\ \citep{Hin2004} and \veritas\
\citep{Hol2006}, which have the sensitivity and angular resolution to
detect these $\gamma$ rays, theoretical progress on this issue would
be particularly timely.

\subsection{Limitations} \label{ssec:discuss-limit}

To bring the discussion to a close, we briefly review the
present limitations of the RFHD approach, and (where possible) suggest
how these limitations might be overcome in future extensions to the
basic approach.

\subsubsection{Cross-field coupling} \label{sssec:discuss-limit-cross}

In deriving the adopted expression~(\ref{eqn:bgrad-approx}) for the
radiative acceleration \bgrad, the polar velocity derivative
$(\partial v/\partial \tmag)_{\rmag}$ is neglected. 
Although the approximation is generally quite reasonable, it necessarily
suppresses any coupling between the flow on adjacent field lines that
may be important in setting the scale of knots and cross-field
discontinuities (\S\ref{ssec:results-2D}).

To include this coupling, we contemplate an extension to the basic
RFHD approach that involves conducting the 1-D simulations lying in
the same meridional plane \emph{in parallel}. By this, we mean that
each numerical timestep advances the flow data of all coplanar field
lines together. This way, the necessary velocity data to calculate
$(\partial v/\partial \rmag)_{\tmag}$ are available throughout the
simulation, and it it not necessary to make the
approximation~(\ref{eqn:dvdtheta-approx}). The additional
computational costs of performing the 1-D simulations in parallel are
quite reasonable; beyond the additional memory overhead, the only
significant issue is that the numerical timestep is limited by the
Courant criterion as applied to all field lines together. This means
that the simulations advance at the rate of the `slowest'
(shortest-Courant time) field line.

\subsubsection{Rigid field ansatz} \label{sssec:discuss-limit-rigid}

A key ingredient of the RFHD approach is the ansatz that field
lines remain rigid.  This applies as long as the magnetic Lorentz
force can balance any competing forces acting perpendicular to field
lines with only minimal distortion from the assumed unstressed
configuration, taken here to be a dipole.  In the wind outflow
regions, the relevant competition can be characterized in terms of the
global wind magnetic confinement parameter, \estar, defined by
\citet{udDOwo2002}.  These authors' MHD simulations show that the
Alfv\'{e}n radius -- where closed field lines become opened into a
radial configuration by the wind -- scales as $\rAlf \approx
\estar^{1/4} \Rstar$.

The parameters adopted for the RFHD simulations
(cf. Table~\ref{tab:params}), together with a polar field strength
$\sim 10^{4}\,{\rm G}$ corresponding to the value inferred for \sorie\
\citep{LanBor1978}, imply a global confinement parameter $\estar
\approx 10^{7}$. The conclusion is therefore that the wind should
substantially distort field lines only for radii $r \gtrsim \rAlf
\approx 50 \Rstar$ -- significantly further out than the maximum shell
parameter $L = 11.2$ considered in our simulations. Although this
simple analysis does not account for the presence of shocks, or for
the effects of the Coriolis force, these additional processes can be
expected to incur modest (order-unity) modifications to the total flow
energy.  Hence, their inclusion should not appreciably modify the
basic \estar-based analysis for the overall competition between field
and flow in regions outside the disk.

However, the radial increase of the {\em centrifugal} force
makes it a potentially important limiting factor in a rigid-field
approach, particularly for the secularly accumulating plasma in the
outer regions of the {\em disk}. In fact, as discussed in the Appendix
of TO05, the centrifugal force acting on this plasma should eventually
become stronger than the available magnetic tension, leading to the
kind of centrifugally driven breakout seen in the MHD simulations of
\citet*{udD2006}.  The timescale for breakout decreases sharply with
local disk radius, asymptotically as $r^{-4}$; moreover, the overall
normalization of this timescale scales with a \emph{disk} confinement
parameter that is quite analogous to, and has a similar magnitude to,
the wind parameter \estar\ (see, e.g., TO05, their eqns. A7 and A8).

In the RFHD simulations, and again adopting the inferred field
strength for \sorie, the outermost, $L=11.2$ field line has a disk
breakout time of $\approx 2\,{\rm Ms}$, roughly an order of magnitude
\emph{shorter} than the simulation duration.  Indeed, over this
duration all field lines having $L \gtrsim 6$ should undergo one or
more breakout episodes, thus formally violating the rigid-field ansatz
when applying the RFHD simulations of these outer regions to \sorie.
But in practice, these regions make only a minor contribution to the
\emph{disk} emission (cf. the lower images of
Fig.~\ref{fig:emission}), implying that the model should still be
applicable for interpreting optical line diagnostics like \Halpha.
Moreover, once field lines reconnect after a breakout event, the wind
shocks should quickly reform to nearly their characteristic asymptotic
strength, meaning that the associated X-ray emission signatures are
also likely to be only weakly affected.  While further direct MHD
simulations would be helpful in testing these expectations, it seems
that models based on the rigid-field ansatz here still provide good
approximate predictions of key observational diagnostics.

\subsubsection{Open field topologies} \label{sssec:discuss-limit-open}

Recall that the general RFHD approach is not necessarily
limited to any particular magnetic topology, such as the dipole form
adopted here.  It merely requires that this topology be pre-specified
and independent of the actual flow.  For example, the basic approach
could even be applied to a topology in which a dipole field is opened
into a radial configuration in the region outside the Alfv\'{e}n
radius.  Physically such opening occurs because of the stresses of
wind outflow, but phenomenologically it can be pre-specified by
invoking current sheets \citep[e.g.,][]{Con1981,Tsy1989,TsyPer1994}
and/or source surfaces \citep[e.g.,][]{AltNew1969} for regions above
\rAlf. In models of the solar wind, the recent investigation by
\citet{Ril2006} confirms that such source-surface methods lead to
global field topologies that are in good agreement with full MHD
solutions.

In the context of massive stars, corresponding open-field RFHD
simulations could provide a first attempt toward a 3-D model for the
effects of the field on open regions of wind outflow.  Although not
contributing appreciably to optical or X-ray emission, the inclusion
of such outflow regions in an RFHD model could prove a good initial
basis for synthesizing phase-dependent UV wind absorption profiles,
for comparison against archival \iue\ spectra.


\section{Summary} \label{sec:summary}

We have presented a new Rigid-Field Hydrodynamics approach to modeling
massive-star magnetospheres. Within the rigid-field ansatz, the
flow along each field line is treated as an independent 1-D
hydrodynamical problem. Using the \vhone\ code, we performed many
separate 1-D simulations of differing field lines, and pieced them
together to build up 3-D model of a \sorie-like star.

These results from these simulations showcase the potential of the
RFHD approach, both by confirming the findings of previous studies
(e..g, wind collision shocks, disc accumulation), and by revealing new
phenomena (e.g., disk oscillations, siphon flows, centrifugal
heating). Hence, we anticipate that the new approach will prove a
powerful tool in future investigations of massive-star magnetospheres.


\section*{Acknowledgments}

We acknowledge support from NASA grants {\it LTSA}/NNG05GC36G and
\chandra/TM7-8002X. We thank David Cohen and Marc Gagn\'{e} for many
interesting discussions throughout the course of this research.


\bibliography{rfhd}

\bibliographystyle{mn2e}


\appendix


\section{The magnetohydrostatic limit} \label{app:static}

The Rigidly Rotating Magnetosphere (RRM) model is based on the
magnetohydrostatic limit, where all velocities and time derivatives in
the Euler equations~(\ref{eqn:hydro-mass}--\ref{eqn:hydro-energy})
vanish. In this limit, the momentum equation becomes
\begin{equation}
\frac{\diff p}{\diff s} = \rho \bgeff \cdot \bes
\end{equation}
(the \bgrad\ term is not present because the radiative acceleration
scales with the velocity gradient; see~\S\ref{ssec:rfhd-rad}). With
the effective gravity expressed in terms of a scalar effective
potential (cf. eqn.~\ref{eqn:geff}), we recover an equation describing
magnetohydrostatic equilibrium along the field line,
\begin{equation}
\frac{\diff p}{\diff s} = - \rho \frac{\diff \poteff}{\diff s}\, .
\end{equation}
Assuming that the stationary plasma has cooled to the stellar
surface temperature \Tstar, the equation of
state~(\ref{eqn:state}) is used to derive the density distribution
\begin{equation} \label{eqn:rho-distribution}
\rho = C \ee^{-\poteff/\csoundstar^{2}}\, ,
\end{equation}
where $C$ is a constant of integration, and \csoundstar\ is the
isothermal sound speed introduced in \S\ref{ssec:code-grid}.

In the vicinity of a local minimum (as sampled along a field line),
the effective potential can be approximated by the second-order Taylor
expansion,
\begin{equation} \label{eqn:poteff-taylor}      
\poteff(s) \approx \poteffmin + \frac{(s - s_{0})^2}{2} \poteffmin''
\, ,
\end{equation} 
where $\smin$ is the arc distance coordinate of the minimum. Here, primes are
used to denote differentiation with respect to $s$, while a subscript
0 indicates evaluation at the minimum $s=\smin$. (The $\poteffmin'$
term in the expansion is by definition zero.) Substituting
this expression into eqn.~(\ref{eqn:rho-distribution}) reveals a
Gaussian density distribution,
\begin{equation} \label{eqn:rrm-rho}
\rho = \rhomin \,\ee^{-(s - \smin)^2/\hmin^{2}}\, ,
\end{equation}
with \rhomin\ a constant based on $C$, and 
\begin{equation} \label{eqn:hmin}
\hmin \equiv \sqrt{\frac{2\csoundstar^{2}}{\poteffmin''}}
\end{equation}
the characteristic scale height.

In the case of a dipole field aligned with the rotation axis (i.e.,
$\beta=0$), the effective potential minima are situated at the
field-line summits, and thus $\smin = 0$. It can then be shown that
\begin{equation} \label{eqn:d2Phi-asymp}
\poteffmin'' \approx \frac{3 G \Mstar}{\rKep^{3}}
\end{equation}
in the limit $L \gg \rKep/\Rpole$ (see TO05, their eqn.~18), where
\begin{equation} \label{eqn:rKep}
\rKep = \left(\frac{G \Mstar}{\Omega^{2}}\right)^{1/3} = \frac{3 \Rpole}{2 w^{2/3}}
\end{equation}
is the Kepler co-rotation radius where the rotation velocity coincides
with the orbital velocity. In the same limit, the disk scale height is
approximated by
\begin{equation} \label{eqn:hinf}
\hmin \approx \hinf \equiv 
\frac{3 \csoundstar}{2 w} \sqrt{\frac{\Rpole^{3}}{G \Mstar}}\, ,
\end{equation}
which depends only on the rotation rate and the stellar parameters. 

\section{The radial velocity derivative} \label{app:velocity}

To evaluate the radial velocity derivative in eqn.
(\ref{eqn:deltav-full}), we first expand it as
\begin{equation}
\left(\frac{\partial v}{\partial \rmag}\right)_{\tmag} =
\left(\frac{\partial v}{\partial s}\right)_{L}
\left(\frac{\partial s}{\partial \rmag}\right)_{\tmag} + 
\left(\frac{\partial v}{\partial L}\right)_{s}
\left(\frac{\partial L}{\partial \rmag}\right)_{\tmag}\, .
\end{equation}
The right-hand side of this expression includes the velocity
gradients both along and across the field. To eliminate the
cross-field gradient $(\partial v/\partial L)_{s}$, we similarly
expand the polar velocity derivative as
\begin{equation}
\left(\frac{\partial v}{\partial \tmag}\right)_{\rmag} =
\left(\frac{\partial v}{\partial s}\right)_{L}
\left(\frac{\partial s}{\partial \tmag}\right)_{\rmag} + 
\left(\frac{\partial v}{\partial L}\right)_{s}
\left(\frac{\partial L}{\partial \tmag}\right)_{\rmag}\, .
\end{equation}
Combining these two expressions, and after some
straightforward algebra, we obtain
\begin{equation}
\left(\frac{\partial v}{\partial \rmag}\right)_{\tmag} =
\left(\frac{\partial v}{\partial s}\right)_{L}
\left(\frac{\partial s}{\partial \rmag}\right)_{L} -
\left(\frac{\partial v}{\partial \tmag}\right)_{\rmag}
\left(\frac{\partial \tmag}{\partial \rmag}\right)_{L}\, .
\end{equation}
This equation is exact; however, under the assumption
(\ref{eqn:dvdtheta-approx}) that the polar velocity
derivative can be neglected, the second term on the right-hand side
vanishes, so that
\begin{equation}
\left(\frac{\partial v}{\partial \rmag}\right)_{\tmag}  =
\left(\frac{\partial v}{\partial s}\right)_{L} \left(\frac{\partial
s}{\partial \rmag}\right)_{L} = 
\left(\frac{\partial v}{\partial s}\right)_{L} \sec\chi 
\, .
\end{equation}
Here, the second equality follows from eqns.~(\ref{eqn:field-r})
and~(\ref{eqn:line-element}), and gives the final
form~(\ref{eqn:dvdr-approx}) for the radial 
velocity derivative.

\section{Linear perturbation analysis} \label{app:perturb}

To investigate small-amplitude departures from the magnetohydrostatic
equilibrium discussed in Appendix~\ref{app:static}, we employ a linear
analysis. Equations~(\ref{eqn:hydro-mass}) and~(\ref{eqn:hydro-mom})
are subjected to small-amplitude disturbances; discarding terms of
quadratic or higher order in the perturbation amplitude leads to the
system of equations
\begin{gather} 
\label{eqn:pert-mass}
\frac{\partial \rhopert}{\partial t} + \frac{1}{A}
\frac{\partial}{\partial s} \left(A \rho \vpert\right) = 0\, , \\
\label{eqn:pert-mom}
\frac{\partial \rho \vpert}{\partial t} + \frac{\partial
  \ppert}{\partial s} = - \rhopert \frac{\diff \poteff}{\diff s}\, .
\end{gather}
Here, the subscript $_{\rm p}$ denotes Eulerian (fixed-position)
perturbations, while quantities without subscripts refer to the equilibrium
state. As before, the \bgrad\ term has been dropped, and the \bgeff\
term is expressed in terms of the effective potential \poteff; no term
containing the unperturbed velocity $v$ appears, because the
equilibrium state is taken to be at rest.

To solve these equations, we make the approximations that (i) all
perturbations share a time dependence of the form $\ee^{\ii \omega
t}$; (ii) the Taylor expansion~(\ref{eqn:poteff-taylor}) may be used
to model the spatial variation of \poteff; (iii) the derivative $\diff
A/\diff s$ can be neglected, and (iv) the perturbations are
isothermal, with the temperature remaining at \Tstar. With these
approximations, the governing equations for the spatial dependence of
\rhopert\ in the vicinity of a potential minimum $s=\smin$ can be
reduced to the single second-order equation,
\begin{equation}
\hmin^{2} \,\rhopert'' + 2(s-\smin) \,\rhopert' +
    \left(\frac{\omega^{2} \hmin^{2}}{\csoundstar^{2}} + 2\right)
    \,\rhopert = 0\, .
\end{equation}
As before, primes denote differentiation with respect to $s$, and
\hmin\ was defined in eqn.~(\ref{eqn:hmin}). Subject to the boundary
conditions that $\rhopert \rightarrow 0$ for $(s-\smin) \rightarrow
\pm \infty$, the eigensolutions are readily found as
\begin{equation}
\rhopert(s) = a_{m} \,\ee^{-(s-\smin)^{2}/\hmin^{2}} \,\Herm[(s-\smin)/\hmin]
\end{equation}
where $a_{m}$ is a constant setting the amplitude of the
perturbations, and \Herm\ is the Hermite polynomial of integer
degree $m \geq 0$. These normal modes have eigenfrequencies set by
the characteristic equation
\begin{equation}
\omega = \omega_{m} \equiv \sqrt{2 m} \frac{\csoundstar}{\hmin}\, .
\end{equation}
The fundamental ($m=0$) mode has a zero frequency, and corresponds not
to an oscillation, but to the addition of more mass to the
disk. Higher order modes all conserve disk surface density [i.e.,
$(\sigdisk)_{\rm p} = 0$], because
\begin{equation}
\int_{-\infty}^{\infty} \ee^{-x^{2}} \Herm(x) \,\diff x = \int_{\infty}^{\infty} \ee^{-x^{2}} \Herm(x) \Hermz(x) \,\diff x = 0
\end{equation}
for $m = 1,2,3,\ldots$; the first equality is because $\Hermz(x) = 1$,
while the second follows from the orthogonality of the Hermite
polynomials with respect to the weighting function $\ee^{-x^{2}}$
\citep[see][]{AbrSte1972}.


\label{lastpage}

\end{document}